%% file: NLG_Benchmarking.tex
\documentclass[12pt]{iopart}

\usepackage{iopams}  
\expandafter\let\csname equation*\endcsname\relax
\expandafter\let\csname endequation*\endcsname\relax
\usepackage{amsmath}

\usepackage{amssymb} 
\usepackage{amsthm}
\usepackage{fullpage} 
\usepackage{graphicx} 
\usepackage{hyperref}
\usepackage[makeroom]{cancel}
\usepackage{tikz} 
\usepackage{tkz-berge}
\usepackage{subcaption}
\usepackage{multirow}
\usepackage{ulem}
\usepackage{bbm}
\usepackage{booktabs}

\newcommand{\cl}[1]{\mathcal{#1}}

\newtheorem{theorem}{Theorem}[section]
\newtheorem{corollary}[theorem]{Corollary}

\newtheorem{prop}[theorem]{Proposition}

\newenvironment{proofof}[1]{\paragraph{Proof#1.}}{\hfill$\square$}
\usepackage{physics}
\usepackage{algorithm}
\usepackage{algpseudocode}
\usepackage{braket}
\usepackage{centernot}

\newcommand{\vect}[1]{\boldsymbol{\mathbf{#1}}}
\usepackage{xcolor}

\usepackage{comment}

\begin{document}

\title{Application-level Benchmarking of Quantum Computers using Nonlocal Game Strategies}

\author{Jim Furches$^1$, Sarah Chehade$^2$, Kathleen Hamilton$^2$, Nathan Wiebe$^{3, 4, 5}$, and Carlos Ortiz Marrero$^{1, 6}$}
\address{$^1$Physical Detection Systems
and Deployment Division, Pacific Northwest National Laboratory,  Richland, WA 99354 }
\address{$^2$Quantum Computational Science Group, Oak Ridge National Laboratory, Oak Ridge, TN 37830}
\address{$^3$Department of Computer Science, University of Toronto, ON M5S 1A1, Canada}
\address{$^4$High Performance Computing Group, Pacific Northwest National Laboratory, Richland, WA 99354}
\address{$^5$Canadian Institute for Advanced Research, Toronto, On M5G 1M1, Canada}
\address{$^6$Department of Computer Science, Colorado State University,  Fort Collins, CO 80523}
 \ead{\mailto{carlos.ortizmarrero@pnnl.gov}}

\begin{abstract}
In a nonlocal game, two noncommunicating players cooperate to convince a referee that they possess a strategy that does not violate the rules of the game. Quantum strategies allow players to optimally win some games by performing joint measurements on a shared entangled state, but computing these strategies can be challenging. We present a variational quantum algorithm to compute quantum strategies for nonlocal games by encoding the rules of a nonlocal game into a Hamiltonian. We show how this algorithm can generate a short-depth optimal quantum strategy for a graph coloring game with a quantum advantage. This quantum strategy is then evaluated on fourteen different quantum hardware platforms to demonstrate its utility as a benchmark. Finally, we discuss potential sources of errors that can explain the observed decreased performance of the executed task and derive an expression for the number of samples required to accurately estimate the win rate in the presence of noise.

\end{abstract}
\maketitle




\section{Introduction}
Running simple instances of quantum algorithms with a provable advantage is difficult given the current state of quantum hardware \cite{dalzell2023quantum, rieffel2024assessing}. For this reason, it is important to develop benchmarking tools and techniques that can test and validate the unique aspects of quantum hardware that are consistent with the predictions of quantum theory. In particular, recent work on quantum benchmarking has highlighted the importance of developing benchmarking metrics that can measure progress toward quantum utility of useful quantum algorithms \cite{proctor2025benchmarking}. 

Low-level benchmark metrics such as randomized benchmarking \cite{emerson2005scalable, knill2008randomized, magesan2012efficient} aim to measure the average error rates of a gate set independent of the initial state or measurement scheme, but are limited, for example, in that it cannot help specify sources of error in an algorithmic pipeline \cite{helsen2022general} and can overestimate gate fidelity in the presence of errors \cite{chen2025randomized}. High-level benchmarks such as Quantum Volume \cite{cross2019validating} aim to measure the performance of the entire quantum computing stack, including all classical control systems, but this can be too broad a metric and does not necessarily capture the performance of useful quantum algorithms. In addition, both of these benchmark metrics are difficult to compute at scale and fail to capture the ability of a specific hardware platform at attaining some quantum advantage. 

Recent work on nonlocal games has begun to shed light into their utility for quantum hardware verification, quantum advantage, and self-testing \cite{natarajan2023bounding, kalai2023quantum, vsupic2020self, hart2024playing, drmota2025experimental}. In a nonlocal game, two noncommunicating players cooperate to convince a referee that they possess a strategy that does not violate the rules of a game. When players are allowed to use entanglement as a resource in the development of their joint strategy, they are able to perform computations that no classical computer can replicate without communication and can win the game with higher probability. Nonlocal games have been historically important and provide a unique setting to explore the relationship between classical physics, quantum theory, and other non-signaling theories~\cite{bell1964einstein, clauser1969proposed,cleve2004consequences,reichardt2012classical}. An extensive body of research links these games to foundational problems in quantum physics, conjectures in operator algebras, and computational complexity theory~\cite{fritz2012tsirelson, slofstra2020tsirelson, ji2021mip}. Moreover, advances in quantum information theory and combinatorics have revealed broad classes of games with a provable quantum advantage when players are allowed to incorporate quantum resources into their strategies, such as graph coloring and graph homomorphism games~\cite{cameron2007quantum, manvcinska2016quantum}, making them exciting experimental candidates for testing quantum hardware~\cite{daniel2022quantum}. Moreover, nonlocal games are classically verifiable, i.e. given a strategy, you can check in polynomial time if the answers satisfy the rules of the game.  

Despite many breakthroughs in our theoretical understanding of nonlocal games, constructing optimal strategies for general nonlocal games remains a challenge. In our work, we propose a new methodology for constructing strategies using variational methods and outline the utility of the strategies found for benchmarking. We begin Section \ref{sec:back} with an introduction to nonlocal games and some definitions. In Section \ref{sec: algo}, we propose the use of a dual-phase optimization technique to find the resource state and the measurement scheme of a quantum strategies for a nonlocal game. In Section \ref{sec: exp}, we demonstrate how our method is able to successfully find optimal strategies for CHSH, an N-Partite Symmetric game, and the graph coloring game. For the graph coloring game, we were able to find a short-depth perfect quantum strategy for a graph on $14$ vertices shown to be the smallest graph instance where there exists a strict separation between classical and quantum strategies \cite{MR2, lalonde2023quantumchromaticnumberssmall}. We then proceed to test the performance of this novel short-depth strategy on $14$ superconducting quantum computing devices and highlight some potential sources of errors causing decreased performance on some of the devices we tested. In Section \ref{sec:theory_practice}, we outline how we can use quantum strategies to benchmark quantum devices, their desirable noise robustness properties, and win rate estimation procedure in the presence of device shot noise. 

\section{Background}\label{sec:back}

\begin{figure}
    \centering
    \includegraphics[width=\linewidth]{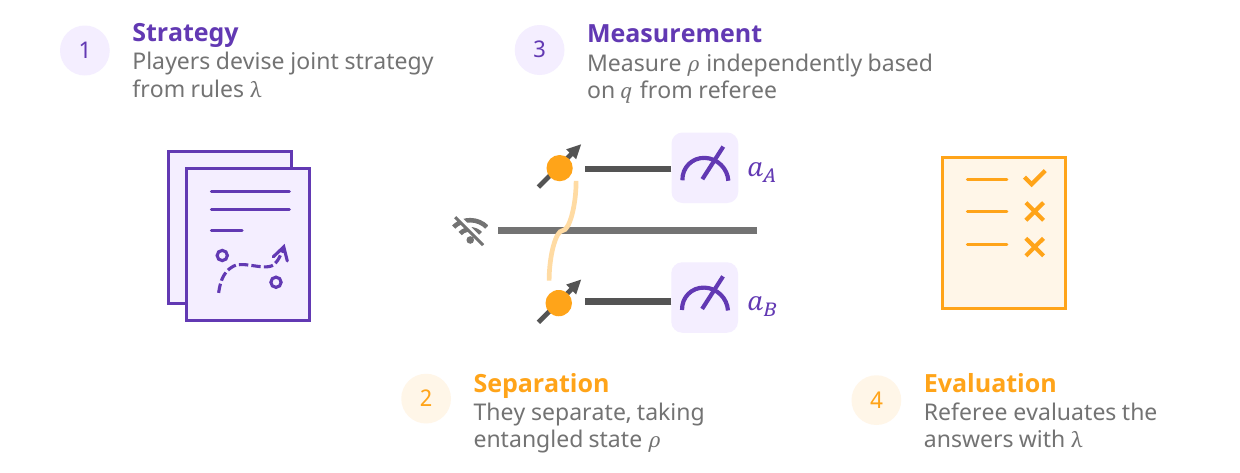}
    \caption{Flow of a nonlocal game. After formulating a strategy, Alice and Bob separate and cannot communicate. For a quantum strategy, they each take a part of an entangled state $\rho$ and upon receiving a question $q$ from the referee, they perform a measurement on their respective states, giving an answer $a \sim p(a|q)$. Finally, the referee receives their answers and verifies them against the rules $\lambda(a|q)$.}
    \label{fig:schematic}
\end{figure}

A nonlocal game of $N$ players $\cl{G} = (Q_1, ..., Q_N, A_1, ..., A_N, \lambda)$ (illustrated in Fig. \ref{fig:schematic}) consists of a set of possible questions $Q_j$ that player $j$ receives from a referee and a set of answers $A_j$ that player $j$ is allowed to send to the referee, which the referee then evaluates against a {\it rule} function $\lambda: Q_1\times ...\times Q_N\times A_1\times ...\times A_N \to \{0,1\}$. Each set of questions $Q_j$ and the set of answers $A_j$ has cardinality $m_j$ and $k_j$, respectively; however, in our work, we assume that there are $m$ questions and $k$ answers for each player. The game proceeds in the following steps:

\begin{enumerate}
    \item The players are informed about the rules of the game $\lambda$, and can collaborate to create a joint strategy, modeled as a conditional probability density between the questions and answers, to maximize their chances of satisfying the rules of the game before it starts.
    \item Players are then separated or isolated to prevent them from communicating. This is referred to as {\it non-signaling}, or in other words, each player's actions are independent of each other.
    \item The referee tests the strategy by asking questions to each player $\mathbf{q} = [q_1, ..., q_N]$ and receiving their responses $\mathbf{a} = [a_1, ..., a_N]$, where $q_i \in Q_i$ and $a_i \in A_i$, respectively.
    \item The players win a round if $\lambda(\mathbf{a}|\mathbf{q}) = 1$, and lose if $\lambda(\mathbf{a}|\mathbf{q}) = 0$. Multiple rounds are played with different questions to establish that players have a valid strategy. 
\end{enumerate}

It is common that all players share the same set of possible questions $Q$ and answers $A$. In particular, {\it Synchronous} games have rules that require that the answers of two (or more) players be identical when asked the same questions $\lambda(a_1, \dots, a_i,a_j,\dots, a_N|\tilde{\mathbf{q}}) = 0$, for all $a_i \neq a_j$, where $\tilde{\mathbf{q}}$ is a vector of questions. In our work, we only consider computing strategies for synchronous games, although the optimization procedure we propose in Section \ref{sec: algo} applies for more general strategies.

Using the rules, we can define the \textit{value} of the game as
\begin{equation}
    \omega(\cl{G}) = \sum_{qa}\lambda(a|q) p(q)p(a|q),
    \label{eq:game_value}
\end{equation}
where the sum is taken over all possible values of $q \in Q^N$ and $a \in A^N$ (we drop the vector notation for convenience). The distribution $p(q)$ of the questions asked is typically chosen to be uniform, and the behavior $p(a|q)$ is determined by the strategy that the players construct. Notice that this is the only term that players can control to maximize their win rate. A strategy is said to be \textit{perfect} if $\lambda(a|q) = 0 \implies p(a|q) = 0$ and, consequently, $\omega(\cl{G}) = 1$.

Classical strategies consist of a lookup table that indexes each player's response to a particular question. It suffices to consider deterministic strategies since stochastic strategies involving shared randomness between the players cannot outperform deterministic strategies due to the linearity of the value of the game \cite{cleve2004consequences}.

Suppose that players share a quantum state $\ket{\psi} \in \otimes_i \cl{H}_i$, and each player has a set of positive operator-valued measures (POVMs) with elements of the form $\{\cl{M}_{a|q}\}$, which they perform on their subspace. Using this setup, players can generate the following conditional probability density,
\begin{equation}
    p(a|q) = \mathrm{Tr}\left[\rho \left(\otimes_i \cl{M}_{a_i|q_i} \right)\right],
\end{equation}
where $\rho=|\psi\rangle\langle\psi|$. 
These densities are known as {\bf quantum strategies}.

In addition to the above definition of a quantum strategy, there are a variety of competing definitions for a quantum strategy depending on the choice of axioms that describe how joint measurements between two parties should be performed \cite{lupini2020perfect}.  In our work, we will only consider strategies as defined above, but the study of quantum strategies is a very active area of research~\cite{helton2021synchronous, helton2019algebras, OP2, manvcinska2021products}. Note that in \cite{cameron2006quantum}, the authors prove that for synchronous games, a maximally entangled state is sufficient for a quantum strategy to win the graph coloring game. 

A game exhibits a {\bf quantum advantage} if there exists a quantum strategy that performs better than the best possible classical strategy, in which case there is a Bell inequality $\cl{I}$ that is violated for some quantum strategies. The inequality has historically served as an experimental test of local realism \cite{chsh}. Such inequalities are constraints satisfied by classical (local hidden-variable) models, and are often linear inequalities derived from the local realism assumption. More specifically, a Bell inequality consists of a function $\cl{I}$ with respect to the probabilities $\{p(a|q)\}$ such that 
\begin{equation}\label{eq:bell_inequality}
    \cl{I}(\{p(a|q)\})\leq \xi,
\end{equation}
for some $\xi\in\mathbb{R}$. 
Bell inequalities are a central object for self-testing of states \cite{vsupic2020self}. The construction of such functions $\cl{I}$ and constants $\xi$ are as follows: for a given Bell inequality $\cl{I}=\sum\limits_{q,a}w_{q,a}p(a|q)$, where $w_{q,a}$ are weights, there is a corresponding Bell operator $\cl{B}=\sum\limits_{q,a}w_{q,a}\bigotimes\limits_{i}\cl{M}_{a_i|q_i}$, such that a violation is obtained as $\xi=\mathrm{Tr}(\cl{B}\rho)$. If the maximal achievable violation is obtained by using quantum resources, denote $\xi_Q$ for this distinction and consider the shifted Bell operator $\xi_Q\mathbbm{1}-\cl{B}$. If the shifted Bell operator admits a decomposition
\begin{equation}
    \xi_Q\mathbbm{1}-\cl{B} = \sum\limits_{\gamma} P_\gamma^\dagger P_\gamma,
\end{equation}
where each $P_\gamma$ is a polynomial with respect to the measurement operators $\{\cl{M}_{a_i|q_i}\}$, then we call the decomposition a \textit{sum of squares} (SOS) for the Bell inequality. Such a decomposition is extremely hard to find \cite{kempe2011entangled}.

\section{Method}\label{sec: algo}

In this section, we present a variational quantum algorithm for computing quantum strategies of nonlocal games. Let $\ket{\psi}$ be the shared entangled state between the players and $\mathcal{M}_{a|q} = \bigotimes_i \mathcal{M}_{a_i|q_i}$ be the joint POVM applied to that state for question $q$, returning $a$ with probability $p(a|q) = \langle \mathcal{M}_{a|q} \rangle$. It was noted in \cite{Bharti_2020} that fixing these measurement operators gives a Hamiltonian whose ground state is the optimal shared state for this measurement setting. This fact has been used with reinforcement learning to optimize measurements while selecting the optimal shared state through exact diagonalization \cite{bharti2019teach}.

\subsection{Dual-Phase Optimization}\label{dual-phase}

Our approach is a dual-phase optimization (DPO) that alternates between 2 phases: preparing the optimal state $\ket{\psi}$ for the fixed measurements $\{\cl{M}_{a|q}\}$, and optimizing the measurements, while fixing the shared state. We assume that the players parameterize their state $\ket{\psi} \rightarrow \ket{\psi(\theta)}$ and POVMs $\cl{M}_{a|q} \rightarrow \cl{M}_{a|q}(\phi)$. The particular choice of parameterization depends on characteristics of the game (e.g. number of qubits required depends on the number of answers).

\begin{algorithm}[!htb]
\caption{Dual-Phase Optimization}
\begin{algorithmic}
\State Initialize $\phi$ randomly
\While {$\Delta \langle H \rangle > \epsilon$}
    \State Construct $H(\phi)$
    \State $\ket{\psi(\theta)} \gets VQE(H(\phi))$
    \State $\phi \gets GD(\langle H(\phi) \rangle)$
\EndWhile
\end{algorithmic}
\end{algorithm}

The preparation of the Hamiltonian depends on the specific measurement scheme the players decide on, which depend on the game. Later, we outline a method for constructing a Hamiltonian from arbitrary game rules $\lambda$ and measurements $\{\cl{M}_{a|q}\}$.

The optimal shared state is prepared in the first phase using any VQE procedure $VQE(\cdot)$. Here, we choose ADAPT-VQE \cite{grimsley2019adaptive} because it generates compact variational circuits for use on near-term quantum hardware, but any other solver can also be used (see \ref{app: vqe}). The reference state $\ket{\psi_{0}}$ can be a product state, e.g. $\ket{0}$, $\ket{+}$. We choose a qubit operator pool consisting of all possible Pauli strings $P$ acting on the entire system. The operators added to the state take the form $e^{i\theta P}$, giving $\ket{\psi(\theta)} = \prod\limits_{j=N}^1 e^{i\theta_j P_j}\ket{\psi_0}$, and they are capable of generating the entanglement required to win nonlocal games, provided they act non-trivially on at least 2 qubits.

The second phase uses a gradient descent-like optimizer $GD(\cdot)$ to update the measurement parameters $\phi$. This requires the calculation of gradients $\nabla_\phi \braket{\psi(\theta)|H(\phi)|\psi(\theta)}$ on the quantum device, which can be done through parameter shift rules \cite{Mitarai_2018, Schuld_2019}. In \ref{app: grad_complexity}, we outline the cost of computing this gradient for larger problem instances and some optimization challenges. 

\subsection{Game Hamiltonians}

As mentioned above, DPO requires the construction of a Hamiltonian based on the measurements of the players, which determines the quantum strategy. Player $i$ may measure their qubits $\rho_i$ in an arbitrary basis depending on the question, leading to a form for the measurement operators
\begin{align}
    \cl{M}_{a|q} &= \bigotimes_i U_{iq_i}^\dagger P_{a_i} U_{iq_i} \\
    &= U_q^\dagger P_a U_q,
\end{align}
where $P_{a_i} = \ketbra{a_i}{a_i}$ is the projector onto answer $a_i$, and $U_{iq_i}$ acts on $\rho_i$ in response to question $q_i$. Because $\langle \mathcal{M}_{a|q} \rangle = p(a|q)$, we can substitute this into (\ref{eq:game_value}) to construct the game operator
\begin{align}
    \beta &= \sum_{qa}\lambda(a|q)p(q)\mathcal{M}_{a|q} \\
    &= \sum_{q} p(q) U_q^\dagger \left(\sum_a \lambda(a|q) P_a \right) U_q
    \label{eq:value_operator}
\end{align}
with the property $\braket{\beta} = \omega(\mathcal{G})$. A VQE finds the ground state of a Hamiltonian, so to maximize the win rate, we use a \textit{value} Hamiltonian $H = -\beta$ in DPO.

\begin{prop}\label{beta}
The value $\braket{\beta} = 1$ if and only if the players have a perfect quantum strategy, otherwise $\braket{\beta} < 1$. 
\end{prop}
\begin{proofof}{~\ref{beta}}
We show this by first computing the value for a perfect strategy and then for an imperfect strategy. Let $I = \{(q,a)~|~\forall q,a~\lambda(a|q) = 0 \}$ be the set of question-answer pairs for which the strategy violates the rules, and let $P = I^c$ be its complement, the set of correctly answered question-answer pairs.

For a perfect quantum strategy, $I = \emptyset$ and $P = Q^N \times A^N$, therefore we get
\begin{align*}
    \braket{\beta} &= \sum_{q,a \in P \cup I} p(q) \lambda(a|q) \braket{\cl{M}_{a|q}} \\
    &= \sum_{qa \in P} p(q,a) + (0)\sum_{qa \in I} p(q,a).
\end{align*}
Since $I = \emptyset$ for all $q,a$ pairs for which $\lambda = 1$ are contained in $P$, it follows that the joint probability density in the left term must sum to 1. Hence, we obtain $\braket{\beta} = 1$.

A very similar line of reasoning holds for an imperfect strategy, where $I \neq \emptyset$. Reusing the above expression,
\begin{align*}
    \braket{\beta} &= \sum_{qa \in P} p(q,a) + (0)\sum_{qa \in I} p(q,a) \\
    &= \sum_{qa \in P} p(q,a) < 1,
\end{align*}
since for $p(q,a),~q,a \in P$ no longer contains the full probability density as $I$ contains some possible pairs. We conclude that $\braket{\beta} \le 1$, with $\braket{\beta} = 1$ iff a strategy is perfect.

\end{proofof}

To parameterize this Hamiltonian for DPO, a general single-qubit unitary $U_1$ may be decomposed into 3 parameters, leading to a parameterization of the full measurement gate $U_q = \bigotimes_{i,j_i} U_1(\phi_{iq_ij_i})$, where $i$ indexes the player, $q_i$ indexes the question, and $j_i$ indexes the particular qubit of player $i$. In measurement layers acting on multiple qubits, we expand each entry of $\phi_{iq_ij_i}$ to be the concatenated vector of parameters for all gates applied to that qubit, i.e. $U_q = \bigotimes_{i} U_{N_i}(\phi_{iq_i})$, where $U_{N_i}$ is an $N_i$-qubit unitary.

\section{Experiments} \label{sec: exp}
To evaluate the performance of DPO, we apply it to several nonlocal games with known quantum bounds: CHSH, the N-partite symmetric (NPS) game \cite{bharti2019teach}, and the odd-cycle game \cite{drmota2025experimental}. Then, we use DPO to explicitly compute an optimal quantum strategy for the coloring game of a $14$ vertex graph called $G_{14}$~\cite{MR2}. This strategy is then evaluated on quantum hardware, demonstrating that it can be used to benchmark the nonlocal capabilities of quantum devices and find sources of errors.

\subsection{CHSH}
The Clauser-Horne-Shimony-Holt (CHSH) scenario \cite{chsh} is the simplest nonlocal game that admits a quantum advantage. CHSH features 2 players, Alice and Bob, who each receive a question $q_i \in Q = \{0, 1\}$, answering $a_i \in A = \{0, 1\}$. The inequality operator can be expressed in the familiar form $\cl{I} = A_0B_0 + A_1B_0 + A_0B_1 - A_1B_1$ following the rules
\begin{equation}
    \lambda(a|q) = 
    \begin{cases}
        a_a \oplus a_b, & \text{if } q_a = q_b = 1 \\
        \delta_{a_a, a_b}, & \text{otherwise}
    \end{cases}
\end{equation}
and making the substitution $\lambda = 0 \rightarrow \lambda = -1$ in (\ref{eq:value_operator}). Here $A_q$ denotes Alice's measurement operator and likewise $B_q$ for Bob. All classical strategies are bounded by $\braket{\cl{I}} \le 2$, whereas quantum strategies can violate this up to $\braket{\cl{I}} = 2\sqrt{2}$. i.e., from Equation (\ref{eq:bell_inequality}), the violation occurs with $\xi_Q=2\sqrt{2}$. It suffices to share a Bell state and then perform the appropriate single-qubit measurements.

We applied DPO to the CHSH game using $R_y(\phi) = e^{-i\frac{\phi}{2}Y}$ gates as the $U_q$ operators. In the first iteration, ADAPT chose $\ket{\psi(\theta)} = e^{i\frac{\pi}{4} YX}\ket{00}$, which correctly generates a Bell state $\ket{\Phi^-}$. In the second phase of that iteration, the measurement parameters were optimized to $\phi \approx [0, -\pi/2, \pi/4, -\pi/4]$ by constraining $\phi_{a0} = 0$, giving the optimal inequality value $2\sqrt{2}$.

\subsection{N-partite Symmetric}

The NPS scenario \cite{nps} involves the correlations between players $N$, each receiving a binary question $q_i \in \{0, 1\}$ and returning a dichotomic answer $a_i = \pm 1$. The inequality is expressed in terms of one- and symmetric two-body correlators,
\begin{equation}
    S_q = \sum_i \braket{\cl{M}_{iq}}, ~ S_{qq'} = \sum_{i \neq j} \braket{\cl{M}_{iq} \cl{M}_{jq'}},
\end{equation}
where measurement $\cl{M}_{iq} = U_{iq}^\dagger Z_i U_{iq}$. The classical bound on the correlations is
\begin{equation}
\cl{I} = -2S_0 + \frac{1}{2}S_{00} - S_{01} + \frac{1}{2}S_{11} + 2N \ge 0,
\end{equation}
with negative values only achievable with quantum strategies \cite{bharti2019teach}.

\begin{figure}[t!]
    \centering
    \includegraphics[width=0.99\linewidth]{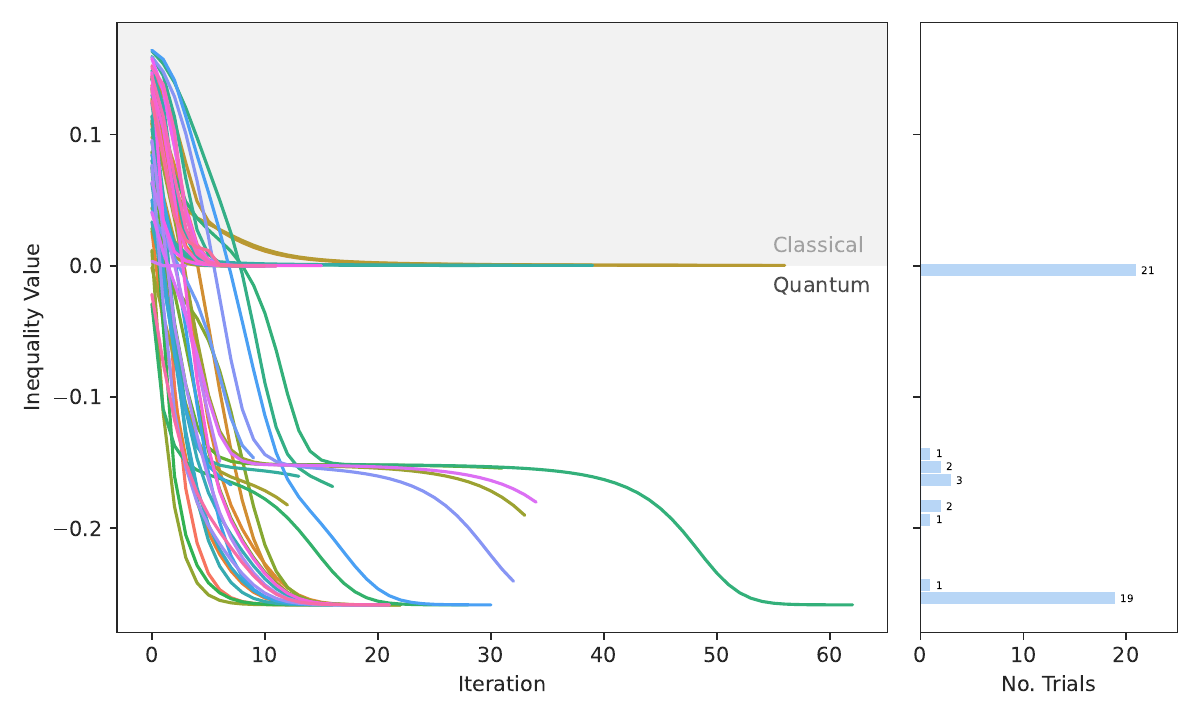}
     \caption{Trials of DPO on NPS for $N=6$. (Left) Trajectory of all 50 trials. Negative inequality values are not reachable with classical states. (Right) Distribution of the final inequality values. Despite the non-convexity of the problem, many trials still reach the optimal value.}
     \label{fig:nps_trials}
\end{figure}

We tested 50 DPO trials for the $N = 6$ case (Fig. \ref{fig:nps_trials}). Our algorithm encounters some local minima, particularly at the classical bound of $\cl{I} = 0$, but still succeeds in 19/50 attempts reaching $\cl{I} = -0.258$ as found in \cite{bharti2019teach} as well. Additionally, 29/50 of the trials violated the classical bound.

\subsection{Chromatic Number Game}\label{sec: coloring}

The objective of the chromatic number game \cite{MR2} is to color a graph $G = (V,E)$ in such a way that adjacent vertices are never given the same color. If this can be done using $c$ colors, we call this a $c$-coloring of the graph. It has been shown recently that winning strategies for this game generate the set of all possible correlations for synchronous nonlocal games \cite{harris2024universality}. This differs from the other nonlocal games we mentioned, as the sets of questions and answers are much larger, and each player requires more qubits to encode their answer. The referee asks a question $q = [v_a, v_b] \in \{(v, v) | v \in V\} \cup E$, and the players respond with colors $a = [c_a, c_b] \in \{1, \dots, c\}^2$. The rules are given by
\begin{equation}
    \lambda(a|q) = \begin{cases}
        \delta_{c_a, c_b} & \text{if } v_a = v_b \\
        (1 - \delta_{c_a, c_b}) & \text{if } (v_a, v_b) \in E
    \end{cases}.
\end{equation}
From these rules, one can encode the graph-coloring game into a Hamiltonian for DPO. For convenience, we denote a vertex question $[v, v]$ as $v$, and an edge question $[v_a, v_b] \in E$ as $e$. Let the answers also be given by $c = [c_a, c_b]$. Then, the expression in (\ref{eq:value_operator}) gives
\begin{equation}
    \beta = \frac{1}{|Q|} \left[\sum_v U_v^\dagger P_{cc} U_v + \sum_e U_e^\dagger (I - P_{cc}) U_v \right],
    \label{eq:graph_coloring_hamiltonian}
\end{equation}
where $P_{cc} = \sum_c \ket{cc}\bra{cc}$ is the projector onto the subspace of matching colors, and $|Q| = |V| + |E|$. Intuitively, the first term maximizes $p(c_a = c_b | v)$, and the second term maximizes $p(c_a \neq c_b | e) = 1 - p(c_a = c_b|e)$. Note that to ensure that all possible questions are asked to each player, $E$ contains both edges $(v_a, v_b)$ and their reverse $(v_b, v_a)$.

First we consider the odd-cycle game \cite{cleve2004consequences}, defined on an odd-length cycle graph $G(n)$ of $n$ vertices. The players are restricted to using 2 colors $c_a, c_b \in \{0,1\}$, meaning there is no perfect classical strategy because $G(n)$ is 3-colorable. Indeed, the optimal classical win rate is $\omega_c(n) = 1 - 1/(2n)$. The optimal quantum strategy has a higher rate of $\omega_q(n) = \cos^2[\pi/(4n)]$, yielding a separation, but no perfect quantum strategy exist. Additionally, this game is of particular interest because it was recently experimentally demonstrated in a spatially separated pair of trapped ions, showing quantum advantage for up to $n \le 19$ vertices \cite{drmota2025experimental}.

\begin{figure}[t!]
    \centering
    \includegraphics[width=0.99\linewidth]{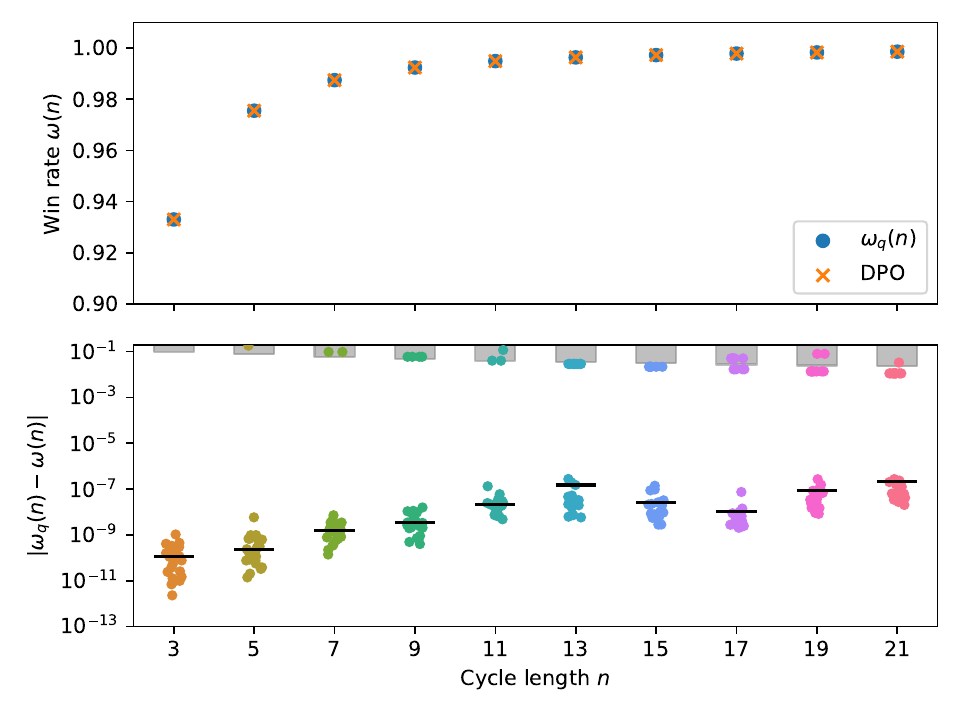}
     \caption{Win rates of discovered strategies for the odd-cycle game. (Top) The maximal win rate found for each game instance $G(n)$ compared to the optimal quantum value $\omega_q(n)$. (Bottom) The distribution of values for each instance. The black lines denote the median, and the gray regions correspond to the classical bound $\omega_c(n)$.}
     \label{fig:odd-cycle-game}
\end{figure}

Fig. \ref{fig:odd-cycle-game} shows the distribution strategies discovered by our algorithm with a measurement layer of one $R_Y$ gate per player. We evaluated each game instance $G(n)$ with 25 trials. In all instances, we observed that the best discovered strategy was within $10^{-8}$ or lower of the optimal quantum value. The algorithm did get stuck in some local minima near the classical value $\omega_c$, but the median values were within the algorithm tolerance, showing that it is able to find graph coloring strategies for the odd-cycle game.

\begin{figure}[h]
    \centering
    \includegraphics[scale=0.3]{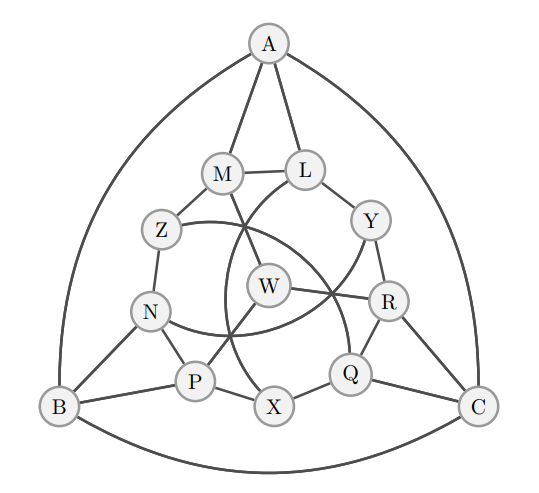}
    \caption{Picture of the $G_{13}$ graph. To get the graph $G_{14}$ just add an apex vertex, vertex connected to all the other vertices, $\alpha$ to this graph. Image courtesy of \cite{MR2}.}
    \label{fig:g13}
\end{figure}

Now we focus on the quantum chromatic game for the graph $G_{14}$ (Fig. \ref{fig:g13}). For this graph there exists a perfect quantum strategy with $4$ colors, while the smallest possible coloring strategy classically requires $5$ \cite{MR2}. Recall that the notion of finding the smallest possible coloring strategy classically is equivalent to finding the chromatic number of this graph \cite{paulsen2015quantum}. This graph was conjectured to be the smallest possible graph with a perfect quantum strategy for this game, and subsequently this was proved to be the case \cite{MR2, lalonde2023quantumchromaticnumberssmall}. In \cite{MR2} a construction was provided using an orthogonal representation of $G_{14}$, that is, a map of vertices to vectors in $\mathbb{R}^4$ such that adjacent vertices are assigned orthogonal vectors. These vectors are then used to define a set of projective measurement operators acting on the maximally entangled state to get a perfect quantum strategy to color the graph. It is unclear how to obtain an explicit set of ans\"atze from the projective measurements to construct a short-depth circuit that can be executed on near-term hardware (see \ref{app:strategy}). We use the DPO algorithm to generate a perfect (up to numerical precision error) quantum strategy for this graph.

To simplify the search for strategies, we restrict the players to 2 qubits each, since 2 qubits suffice to represent $c \in \{0,...,\chi_q - 1\}$ using a binary encoding. We also impose a known constraint on an optimal strategy for synchronous games \cite{MR2, helton2021synchronous}: Bob's measurement operators are complex conjugate to Alice's, halving the number of measurement parameters $\phi$ required. We use a measurement layer per player of general single-qubit $U$ gates, a CNOT from qubit 0 to qubit 1, and $R_y$ gates applied to each qubit, resulting in 8 parameters per question or 112 in total (in our code, this is the \texttt{U3Ry} layer).

\begin{figure}[h]
    \centering
    \includegraphics[width=0.75\linewidth]{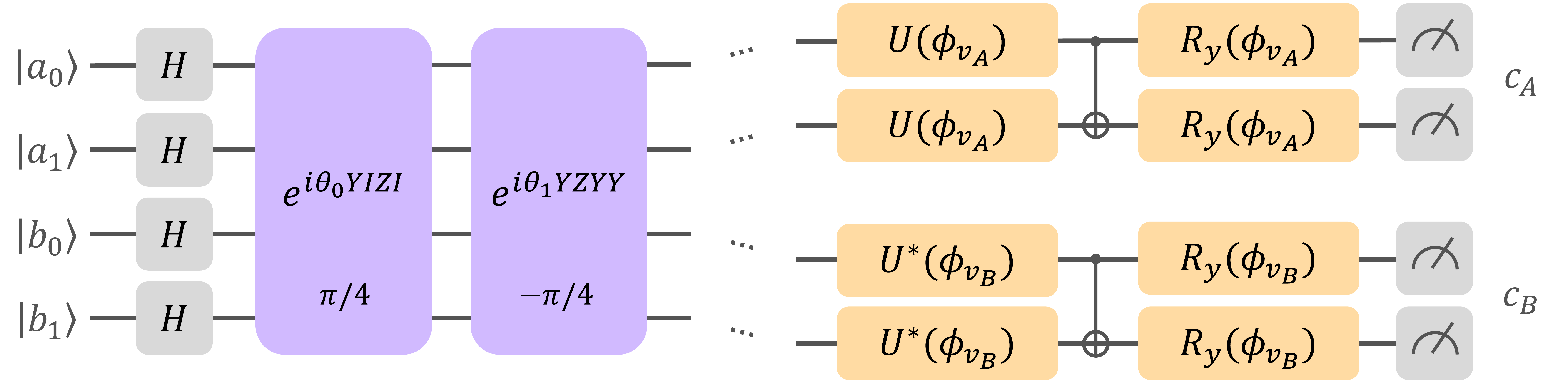}
    \caption{Generated circuit for $G_{14}$. The initial state $\ket{+}^{\otimes N}$ is prepared, then ADAPT added the operators $Y_0 Z_2$ and $Y_0 Z_1 Y_2 Y_3$, giving the shared state $\ket{\psi(\theta)}$. The remaining gate layers along with the 112 measurement parameters $\phi$ constitute the measurement strategy. We only adaptively added circuits on the state preparation and fixed the measurement scheme in this case.}
    \label{fig:circuit}
\end{figure}

We classically simulated 500 trials of DPO, achieving a minimum energy of $E = -1.0000$. We remove the gates added by ADAPT with $|\theta_i| < 10^{-4}$. The corresponding circuit was then converted into a Qiskit circuit (Fig. \ref{fig:circuit}), and the evaluation using the classical \texttt{AerSimulator} confirmed a 100\% win rate (Fig. \ref{fig:g14_error_aer}). The 112 parameters $\phi$ can be found in our code repository (see \ref{app: data}).

It is worth noting that in a nonlocal game the referee cannot cross-check answers from previous questions (otherwise the graph would not be 4-quantum colorable), and the players change their coloring for each vertex probabilistically in subsequent runs, using the entanglement to coordinate their answers as required. For example, when asked $q = [A, A]$ multiple times, the responses are nearly uniform among $[\chi_q]$ but always match. Furthermore, we found that measurement layers consisting of only single-qubit gates were insufficient and generated imperfect strategies at $E = -0.9921$. In these cases, we frequently observed that the errors consisted of a cyclic path through some graph edges.

 The operators chosen by ADAPT are nonlocal as expected, acting on 2 and 4 qubits. The shared state discovered,
\begin{align}
    \ket{\psi(\mathbf{\theta})} &= e^{-i\frac{\pi}{4} YZYY} e^{i\frac{\pi}{4} YIZI}\ket{+}^{\otimes 4} \\
    &= \frac{1}{2}H^{\otimes 4}\sum_{c \in [\chi_q]} \ket{cc},
    \label{eq:shared_state}
\end{align}
 is the maximally entangled state followed by 
 local Hadamard gates. This matches the existing strategy described in \cite{MR2}, which leverages the maximally entangled state, up to local unitary operations.

\begin{figure}[h]
  \centering
  \includegraphics[width=0.75\textwidth]{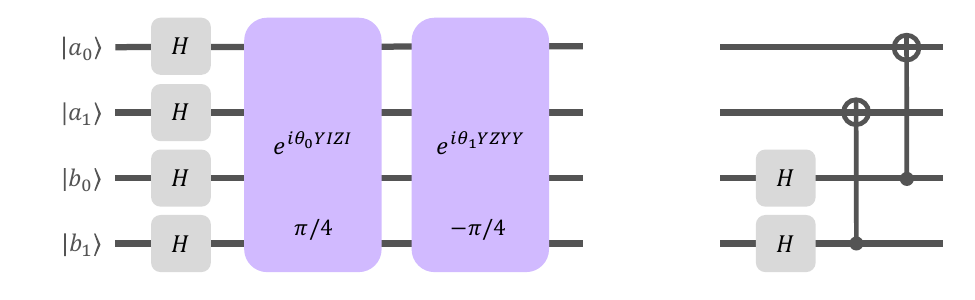}
  \caption{Simplification of the $G_{14}$ state preparation. (Left) The original strategy using the gates found with the adaptive procedure. (Right) The ``Bell pair'' strategy, an equivalent circuit producing two independent Bell state pairs between the players. This requires just 2 CNOT gates compared to the 8 required for the original strategy.}
  \label{fig:bell_pair_strategy}
\end{figure}

 The circuit preparing the shared state $\ket{\psi}$ needs 8 CNOT gates to be transpiled using a ladder-like formulation with CNOT gates applied between nearest-neighbor qubits. This can be reduced to 2 CNOT gates (Fig. \ref{fig:bell_pair_strategy}) by noting that the state $\frac{1}{2}\sum_c \ket{cc}$ can be generated with transversal Bell pairs shared between the players on qubits $a_0, b_0$ and $a_1, b_1$. Applying the transversal Hadamards $H^{\otimes 4}$ in (\ref{eq:shared_state}) flips the direction of the CNOT gates using a simple circuit identity. We refer to this version of the shared state circuit as the ``Bell pair'' strategy, which uses the same measurement layer and parameters as the original strategy.

\subsection{Experiments on IBM Devices}
\label{sec:hw_results}

\begin{figure}
    \centering
    \begin{subfigure}[t]{0.4\textwidth}
    \centering
        \includegraphics[width=1\linewidth]{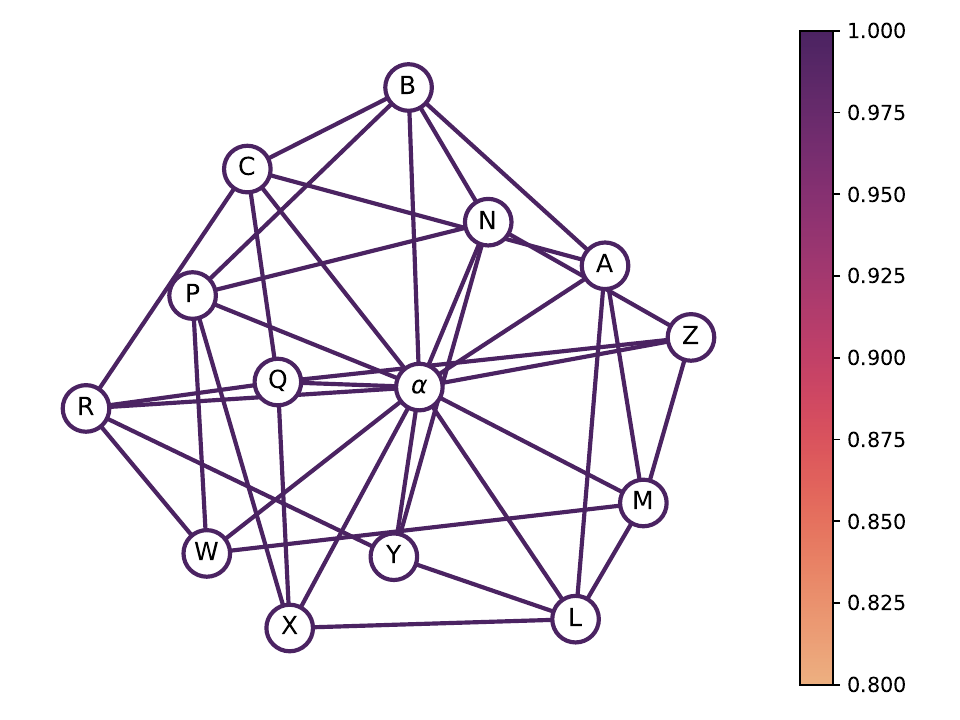}
        \caption{Qiskit \texttt{AerSimulator}}
        \label{fig:g14_error_aer}
    \end{subfigure}
    ~
    \begin{subfigure}[t]{0.4\textwidth}
    \centering
        \includegraphics[width=1\linewidth]{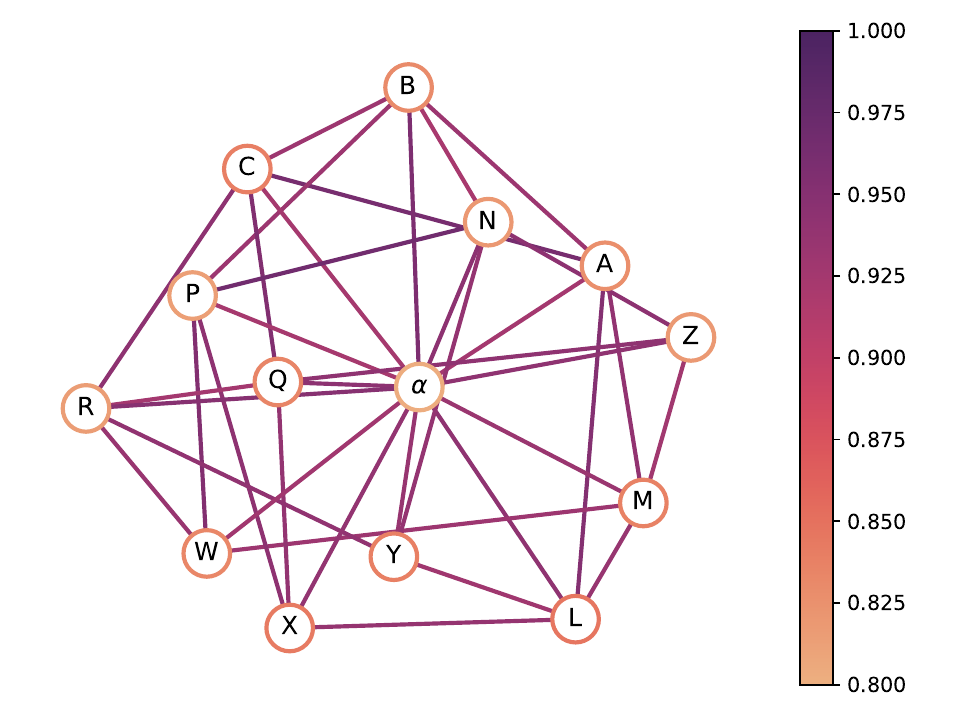}
        \caption{IBMQ \texttt{ibm\_hanoi}}
        \label{fig:g14_error_hanoi}
    \end{subfigure}
    \caption{Win rate of the original $G_{14}$ strategy by question. \subref{fig:g14_error_aer}) Classical statevector simulation with 10000 shots per question. \subref{fig:g14_error_hanoi}) Performance on a quantum device with 1024 shots per question.}
    \label{fig:g14_error}
\end{figure}

\begin{figure}
    \centering
    \begin{subfigure}{0.45\linewidth}
        \centering
        \includegraphics[width=1\linewidth]{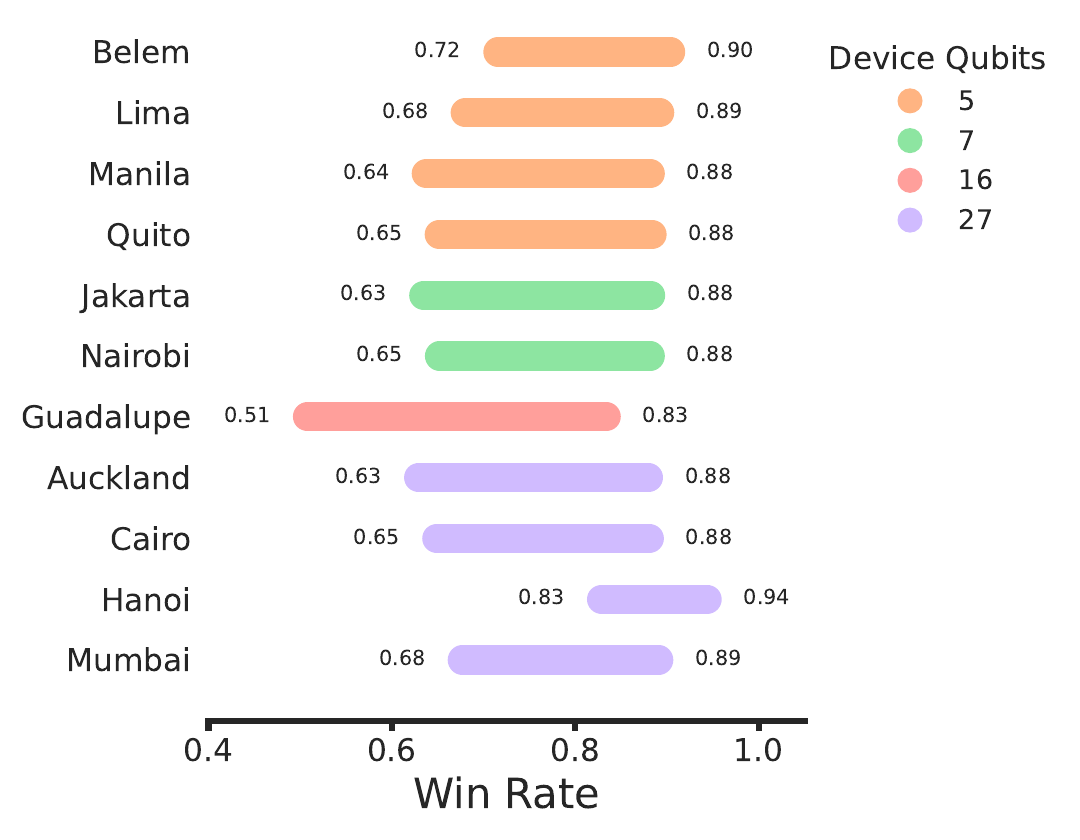}
        \caption{All devices}
        \label{fig:all_ibm_devices}
    \end{subfigure}
    ~
    \begin{subfigure}{0.45\linewidth}
        \centering
        \includegraphics[width=1\linewidth]{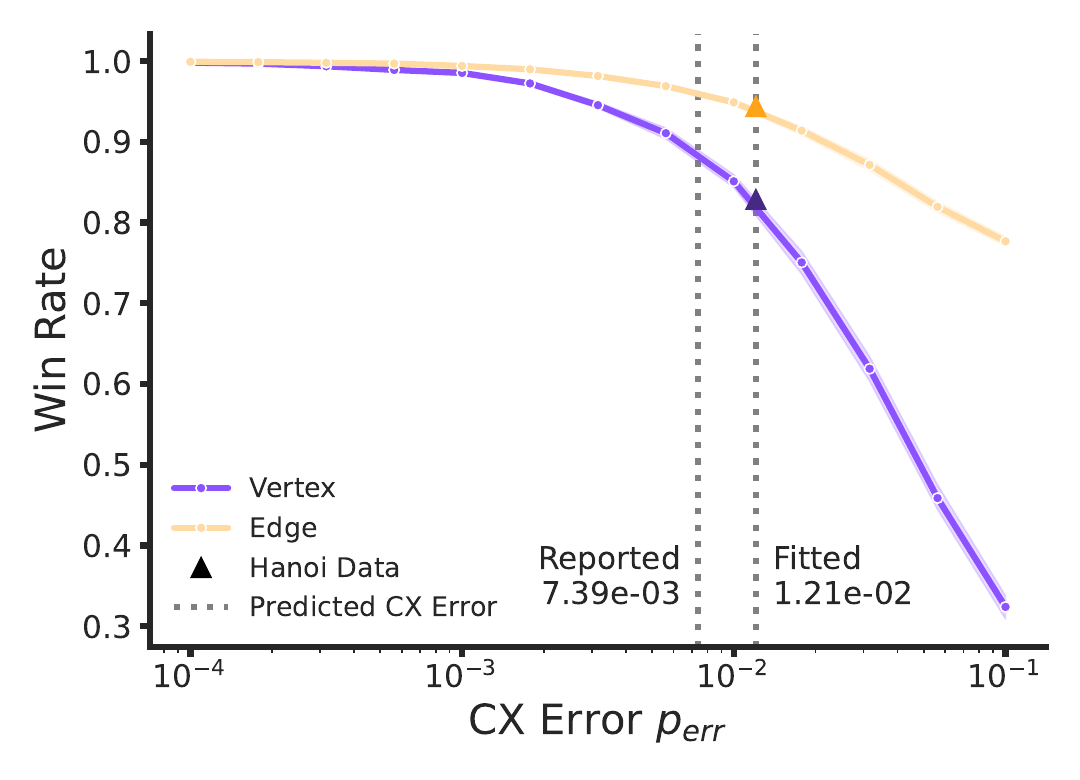}
        \caption{Noise Simulation}
        \label{fig:g14_noise_simulation}
    \end{subfigure}
    \caption{\subref{fig:all_ibm_devices}) Average performance of the strategy on quantum devices grouped by question category, either a vertex $q = [v, v]$ or an edge $q = [v_1, v_2]$ (vertex winrate on the left, edge winrate on the right). The number of device qubits is reported to distinguish processor types; the circuit was executed on 4 qubits. \subref{fig:g14_noise_simulation}) Classical simulation of the circuit with random Pauli noise applied to CNOT gates with probability $p_{err}$. The circuit was transpiled to the basis gates and coupling map of \texttt{ibm\_hanoi}, and the observed data are fitted to the curves by maximizing the probability of observing the data assuming a normal distribution. The estimated error for CNOT gates is higher than reported, since our simulation did not account for measurement readout or decoherence errors.}
    \label{fig:error_byquestion}
\end{figure}

This strategy was submitted to $11$ IBM quantum devices with $4$ or more qubits (Fig. \ref{fig:g14_error}). A decrease in performance was observed on IBM quantum devices compared to the classical simulation due to noise, particularly affecting the success rate of vertex questions (Fig. \ref{fig:g14_error_hanoi}, \ref{fig:error_byquestion}). There are several possible sources of noise:

\begin{enumerate}
    \item Vertex questions are more sensitive to bit flips, as any 1-bit error will result in the answer violating the rules $X_j\ket{cc} \rightarrow \ket{c_ac_b}$, while the same is not true for edge questions, since bit flips may not necessarily make the answers agree $X_j\ket{c_ac_b} \centernot\rightarrow \ket{cc}$ (see Section \ref{sec:sample_complexity}). This asymmetry comes from the rules of the game.

    \item As the resource state depends on entanglement, error in entangling 2-qubit gates disrupts the strategy.

    \item Circuit transpilation to hardware with fixed qubit connectivity further incurs two-qubit gate overhead.
\end{enumerate}
 
 This sensitivity suggests that measuring the win rate of the strategy for $G_{14}$ is a good benchmark to evaluate the ability of a quantum device at accurately controlling for bit flip errors, while simultaneously performing nonlocal operations. In particular, the vertex question win rate is very sensitive to noise, measuring the fidelity of the device gates, whereas the edge question win rate can confirm if a device is using quantum resources. The optimal classical strategy of 4 colors consists of a 4-coloring of $G_{13}$ and assigning the most infrequently used color to the apex vertex $\alpha$. Therefore, all vertex questions would be correctly answered and one edge would be incorrectly answered, resulting in an edge win rate of $36/37 \approx 97.3\%$, or an overall win rate of $86/88 \approx 97.7\%$. Any win rate higher than this requires quantum resources. In our experiments, no device exceeded this threshold (Fig. \ref{fig:error_byquestion}).  However, introducing an error-correcting version of our quantum strategy could improve the robustness of this test, which we leave to future investigations.

\section{Nonlocal Games as Quantum Hardware Benchmarks}
\label{sec:theory_practice}
Nonlocal games with perfect strategies can serve as hardware benchmarks by assessing and analyzing the empirical win rates when executed on near-term hardware.  Under certain assumptions about the structure of quantum noise, nonlocal games can exhibit quantum advantage in shallow circuits, even with noisy qubits \cite{bravyi2020quantum}. The `noisy entanglement' generated in shallow circuits enables correlations that classical circuits fundamentally struggle to reproduce. This is seen in \cite{bravyi2020quantum}: their classical circuits of constant depth cannot simulate the long-range correlations.

In this section, we demonstrate the effectiveness of this benchmark by analyzing hardware noise and its strong correlation to strategy performance. We proceed backwards,from the unentangled readout measurements, to the independent Bell state measurements, to the initial entangled resource state preparation. By investigating the effects of hardware noise on the empirical win rates we seek to establish: a) which questions are most affected by noise, b) which components of the circuit are most affected by noise, and finally, if classical correlations, or quantum noise, could be misinterpreted as a winning quantum strategy. 

\begin{figure}[!htb]
  \centering
  \includegraphics[width=\textwidth]{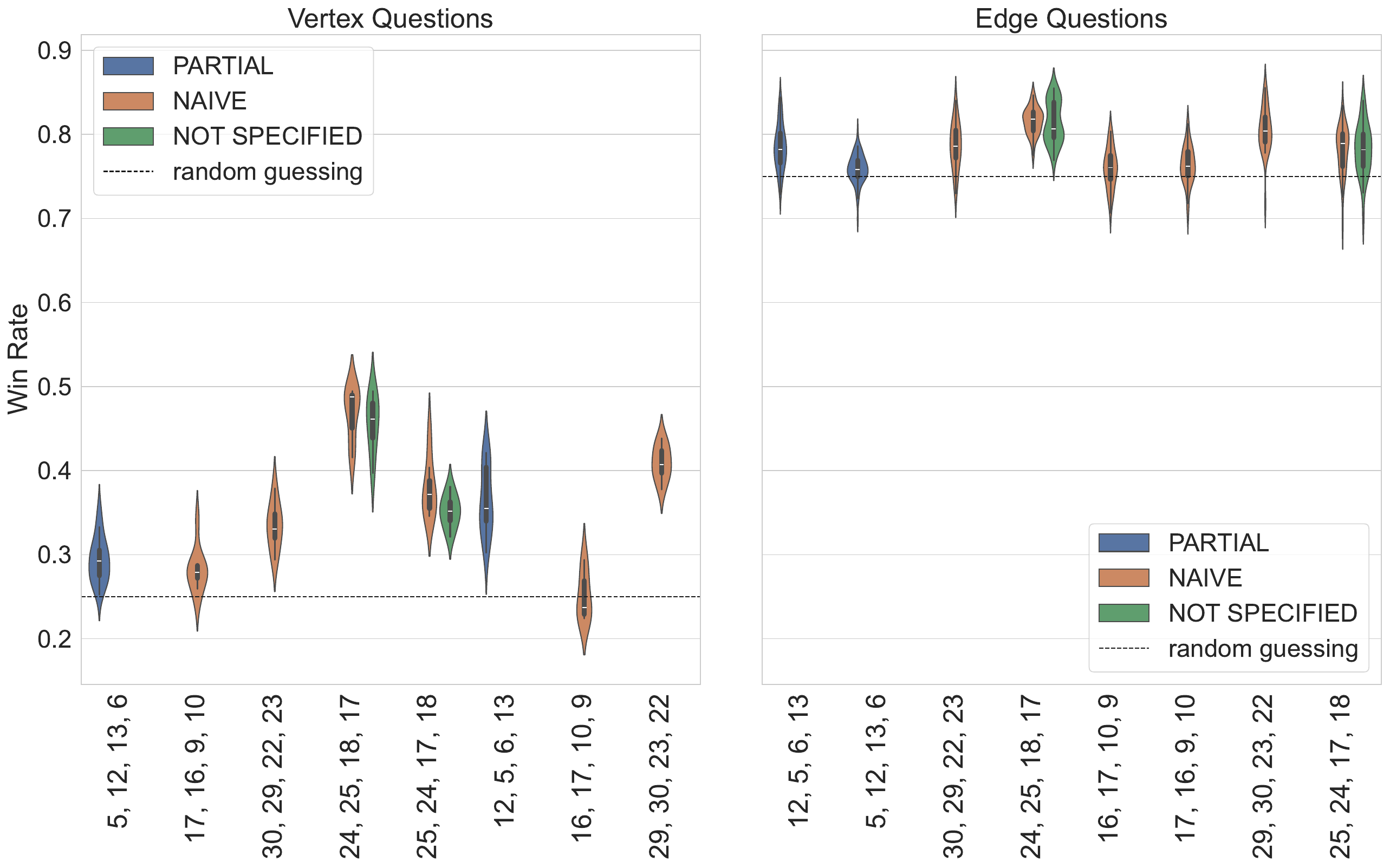}
  \caption{(Left) All vertex question win rates of the original $G_{14}$ strategy grouped by hardware qubits used on \texttt{Ankaa-2} from Rigetti. (Right) Edge question win rate of the original $G_{14}$ strategy grouped by hardware qubits used on \texttt{Ankaa-3} from Rigetti.}
  \label{fig:ankaa2_G14}
\end{figure}

\begin{figure}[!htb]
  \centering
  \includegraphics[width=0.75\textwidth]{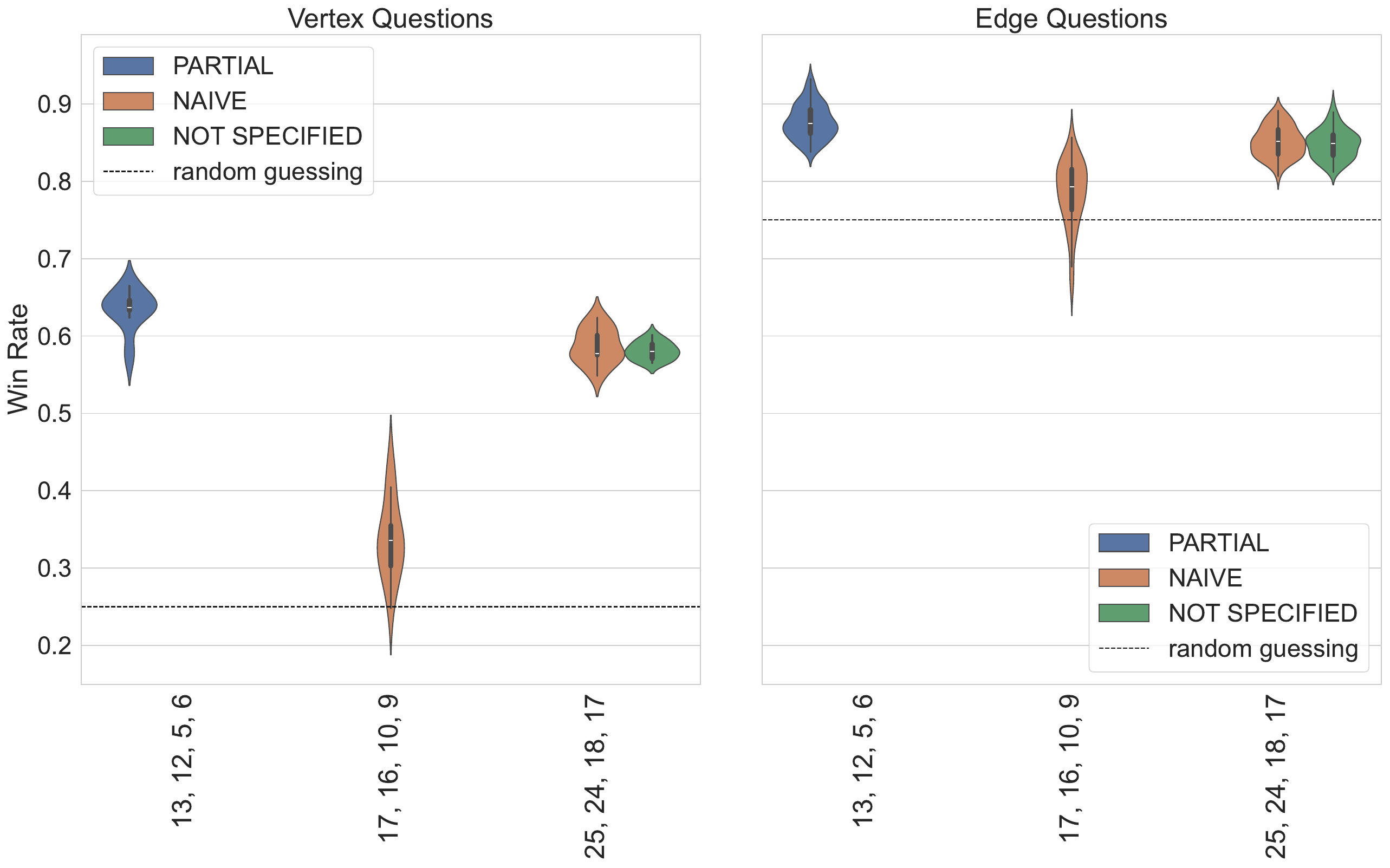}
  \caption{(Left) Vertex question win rate using the Bell pair strategy averaged over multiple experiments on Rigetti's \texttt{Ankaa-2}. The runs used different hardware qubits and wiring strategies. (Right) Edge question win rate using the Bell pair strategy averaged over multiple experiments on Rigetti's \texttt{Ankaa-3}.}
  \label{fig:ankaa2_bell_pair}
\end{figure}

\begin{figure}[!htb]
  \centering
  \includegraphics[width=\textwidth]{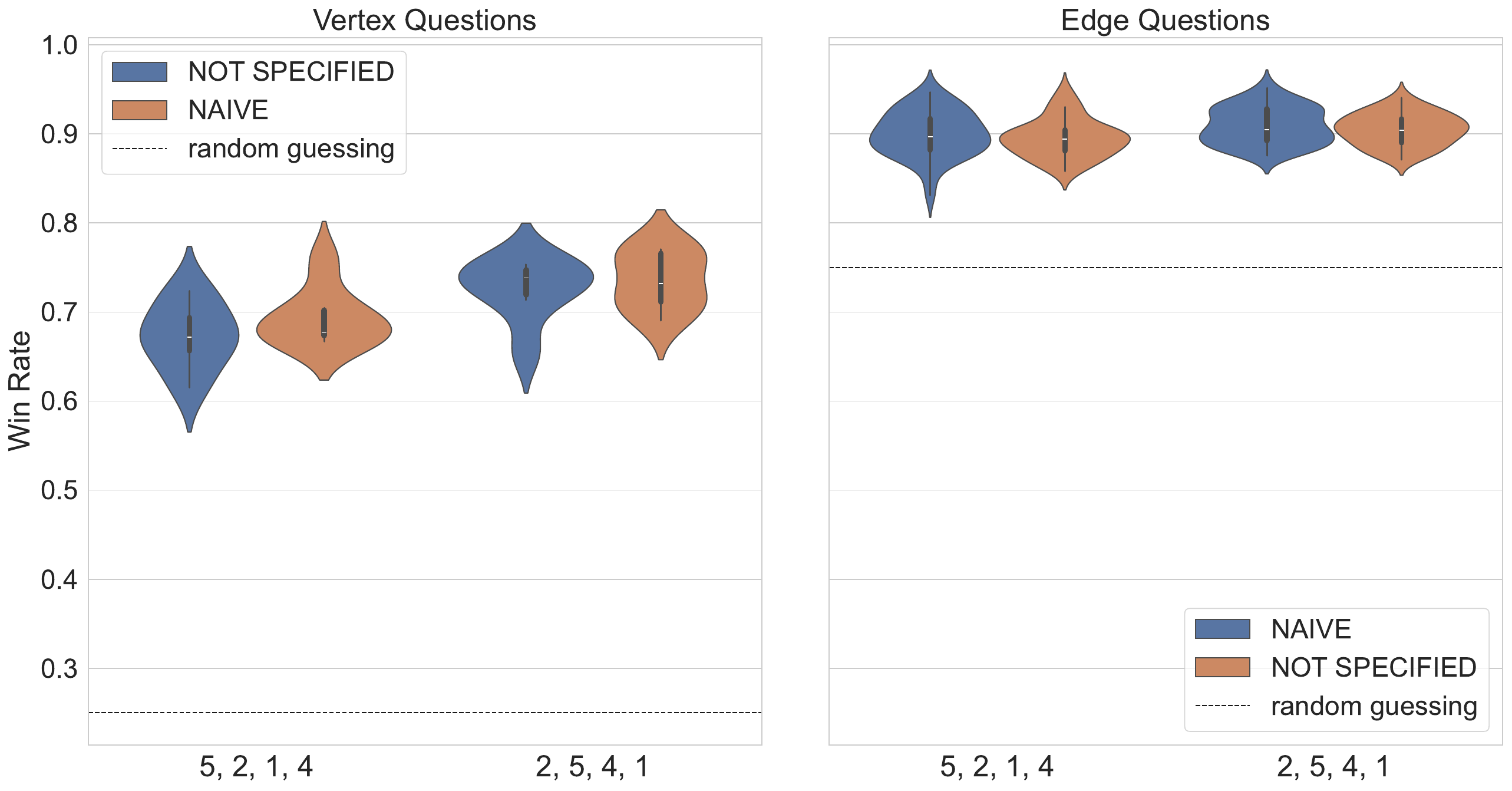}
  \caption{(Left) All vertex question win rates of the original $G_{14}$ strategy grouped by hardware qubits used on \texttt{Ankaa-3} from Rigetti. (Right) Edge question win rate of the original $G_{14}$ strategy grouped by hardware qubits used on \texttt{Ankaa-3} from Rigetti.}
  \label{fig:ankaa3_G14}
\end{figure}

\begin{figure}[!htb]
  \centering
  \includegraphics[width=0.75\textwidth]{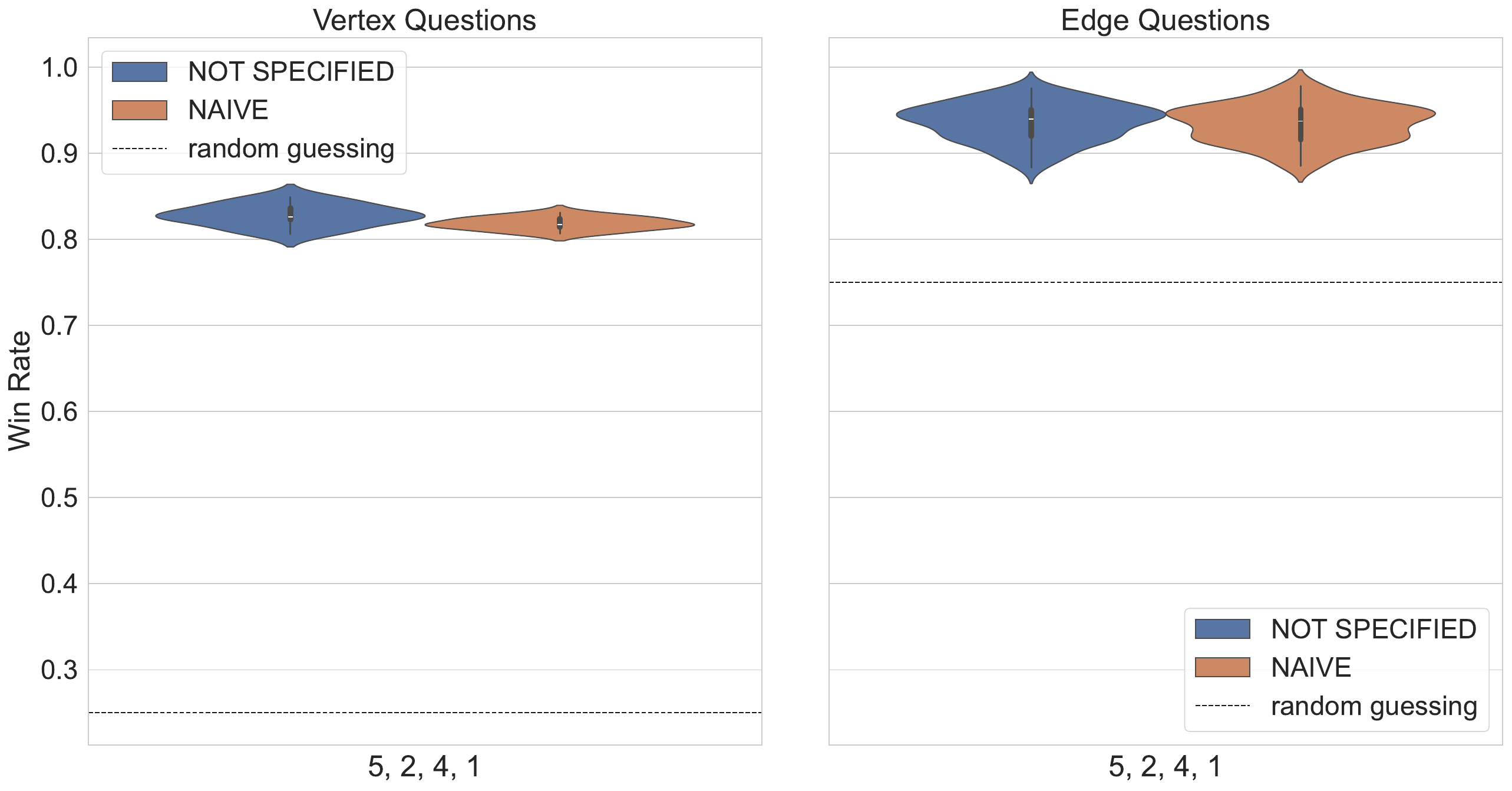}
  \caption{(Left) Vertex question win rate using the Bell pair strategy averaged over multiple experiments on Rigetti's \texttt{Ankaa-3}. The runs used different hardware qubits and wiring strategies. (Right) Edge question win rate using the Bell pair strategy averaged over multiple experiments on Rigetti's \texttt{Ankaa-3}.}
  \label{fig:ankaa3_bell_pair}
\end{figure}

In addition to the classical bounds provided in Section \ref{sec:hw_results}, we also consider the worst outcome on hardware: a nearly uniform distribution over all bitstrings.  This would skew the win rates in the $G_{14}$ game as follows: for any vertex question would only be $1/4 = 0.25$ and the average win rate of any edge question would be $12/16 = 0.75$. Thus random guessing would return an overall win rate of 59\%. In Figures \ref{fig:ankaa2_G14}, \ref{fig:ankaa2_bell_pair}, \ref{fig:ankaa3_G14} and \ref{fig:ankaa3_bell_pair} we include these values as a reference.  

Quantum hardware is affected by many sources of noise. The noise profile is time dependent and there are many strategies developed to optimize the scheduling and execution of quantum circuits. Using superconducting qubit platforms from IBM and Rigetti, we investigate the robustness of the original $G_{14}$ strategy, and the Bell pair strategy on superconducting qubit platforms with respect to hardware noise fluctuations over several days, and also to changes in the circuit made during the transpiration step.

\subsection{Theoretical Noise Robustness}
\label{sec:hamming_bound}
There is an asymmetric sensitivity to noise between the vertex and edge questions due to the game rules (see Section \ref{sec:hw_results}). Furthermore, there is variance in the edge questions performance.  We hypothesize this arises from the particular strategy and distribution of answers found via the ADAPT algorithm. We further investigate the effects of bit-flip errors on the game strategies.

Multiple bitstrings satisfy the constraints for edge questions $\lambda(c, c|v_A, v_B)=0$ for all colors $c$. Players using the four qubit strategy can win an edge question by outputting a bit string that is either Hamming distance $H(a,b)=1$ from matching (e.g. 0001) or distance 2 (e.g. 1100). While both options satisfy the game rules, choosing bitstrings with a greater Hamming distance reduces the likelihood of losing due to device noise, as higher-weight errors occur less frequently.

\begin{figure}[h]
  \centering
  \includegraphics[width=0.75\textwidth]{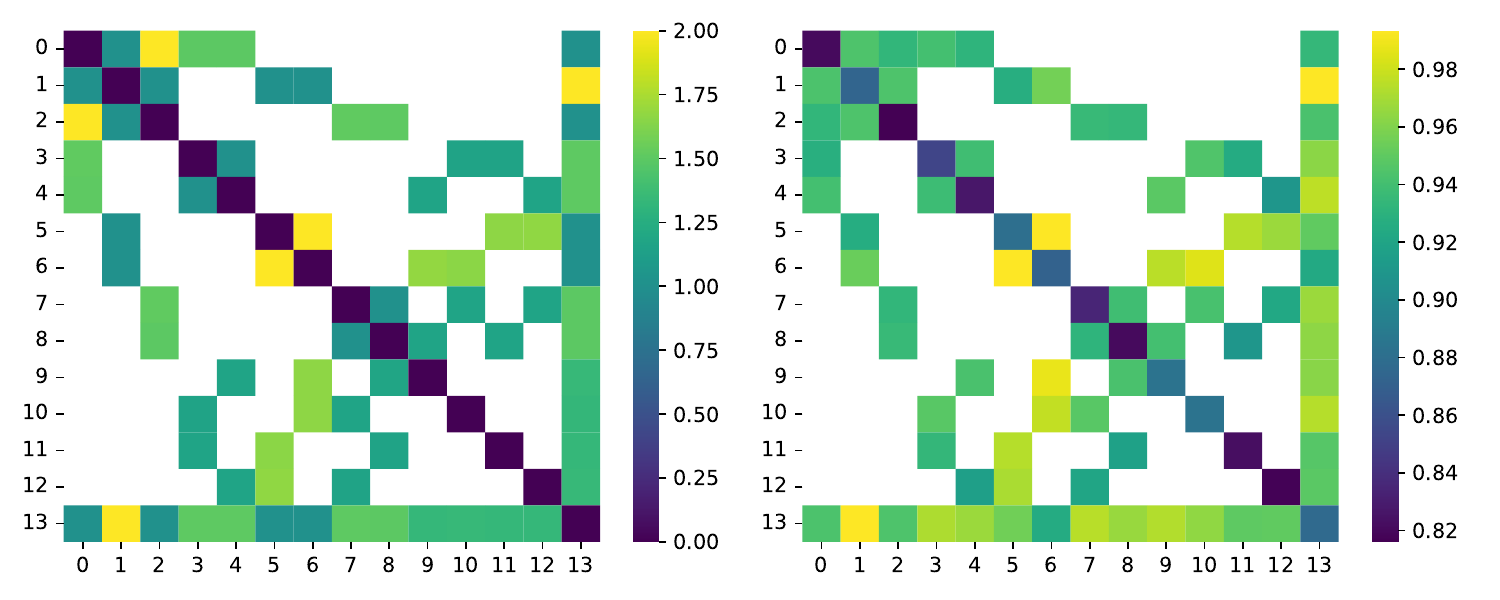}
  \caption{(Left) Noise robustness of the $G_{14}$ strategy. The adjacency matrix of the $G_{14}$ graph is shown with the color scale denoting the EHD. Vertex questions form the main diagonal. (Right) Average performance on \texttt{ibm\_sherbrooke} over 7 runs plotted for comparison (same as in Fig. \ref{fig:sherbrooke_mean_and_std}).}
  \label{fig:expected_hamming_distance}
\end{figure}

To quantify the noise robustness of the strategy resulting from this, we introduce the expected Hamming distance (EHD),
\begin{equation}
    EHD(v_a, v_b) = \mathbb{E}_{c_a, c_b \sim p(c_a, c_b | v_a, v_b)} \left[H(c_a, c_b)\right],
    \label{eq:expected_vertex_distance}
\end{equation}

where $H(a, b)$ denotes the Hamming distance between the binary representations of answers $a$ and $b$. In general, the EHD is not efficiently computable on a classical computer since it requires sampling the strategy. However, because the $G_{14}$ strategy is sufficiently small, we calculate the EHD for each circuit via classical simulation (Fig. \ref{fig:expected_hamming_distance}).

\begin{figure}[h]
  \centering
  \includegraphics[width=0.75\textwidth]{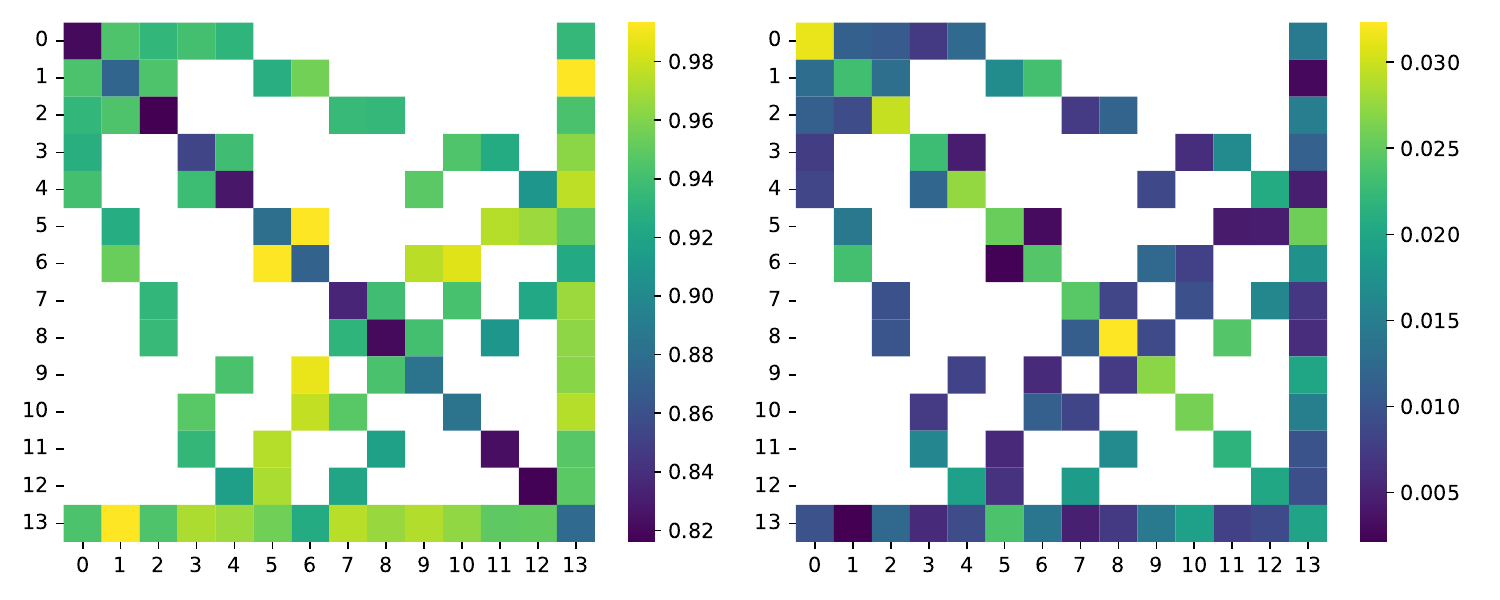}
  \caption{(Left) Win rate of the Bell pair strategy averaged over 7 separate experiments on \texttt{ibm\_sherbrooke}. All runs used the same hardware qubits and layout, forming a linear chain. Because the experiments were spaced out, the calibration parameters differed between each trial. (Right) Standard deviation of each question for those experiments. This shows how sensitive each circuit in the strategy is to fluctuations in the device parameters.}
  \label{fig:sherbrooke_mean_and_std}
\end{figure}

To show that the EHD effectively predicts question performance, we also plot results collected on \texttt{ibm\_sherbrooke}\footnote{Eagle r3 processor}, a superconducting qubit platform available from IBM with 127 qubits. We executed the strategy described in Section \ref{sec: coloring} multiple times over the course of a week. Supplemental experimental details are available in the \ref{app: experimental_details}. The heatmaps exhibit a high degree of correlation ($r = 0.8812, p < 0.001$), suggesting the strategy produced greatly influences noise robustness. The standard deviation is also presented alongside the win rate (Fig. \ref{fig:sherbrooke_mean_and_std}), further highlighting the sensitivity of different questions to variations in device calibration. There are some outliers, particularly $(0, 2)$ and $(2, 0)$ that perform worse than expected, and the EHD cannot account for variation in the vertex question performance. We leave these to future investigations.

We executed circuits for the original $G_{14}$ strategy and the Bell pair strategy on Rigetti's \texttt{Ankaa-2}, and \texttt{Ankaa-3} devices. Both have square lattice qubit connectivity and to take advantage of this, we prioritized running experiments on qubit subsets with cyclic connectivity.  The circuits first constructed in Qiskit are exported to Open Quantum Assembly Language (QASM) \cite{cross2017open}, then imported and compiled into Quil using the \texttt{qiskit-rigetti} plugin \cite{qiskit-rigetti}. During compilation into native operations that are executable on the Rigetti platforms, circuit optimization is possible.

The compiled circuit can be further optimized through rewiring directives that determine how program qubits are mapped onto hardware qubits.  The NAIVE rewiring uses the program qubit register index as the hardware qubit index.  This rewiring may require the use of additional operations to mitigate non-neighboring interactions.  The PARTIAL rewiring attempts to optimize the mapping between program and physical qubits to optimize the fidelity of the compiled circuit. We specified the rewiring strategy through the use of pre-compilation hooks.  If no hooks were specified by the user, the rewiring strategy was not verified and we denote the strategy as (NOT SPECIFIED).  

\subsection{Noise Robustness of Game Components}
\label{sec:noise_characterization}
\begin{figure}
    \centering
    \begin{subfigure}{0.45\linewidth}
        \centering
        \includegraphics[width=1\linewidth]{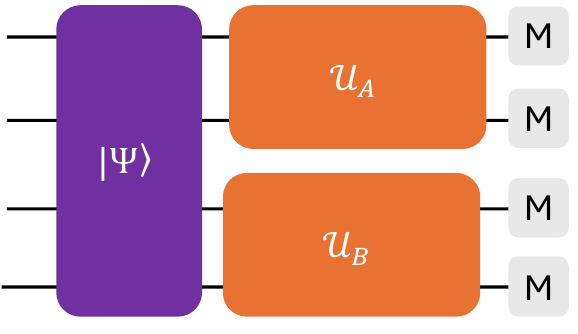}
    \end{subfigure}
    \caption{The components of the nonlocal game circuit that we characterize: resource state (purple), independent Bell basis measurements (orange), and readout error (grey).}
    \label{fig:noise_char}
\end{figure}

\begin{figure}
    \centering
     \centering
    \begin{subfigure}[b]{\linewidth}
        \centering
        \includegraphics[width=\linewidth]{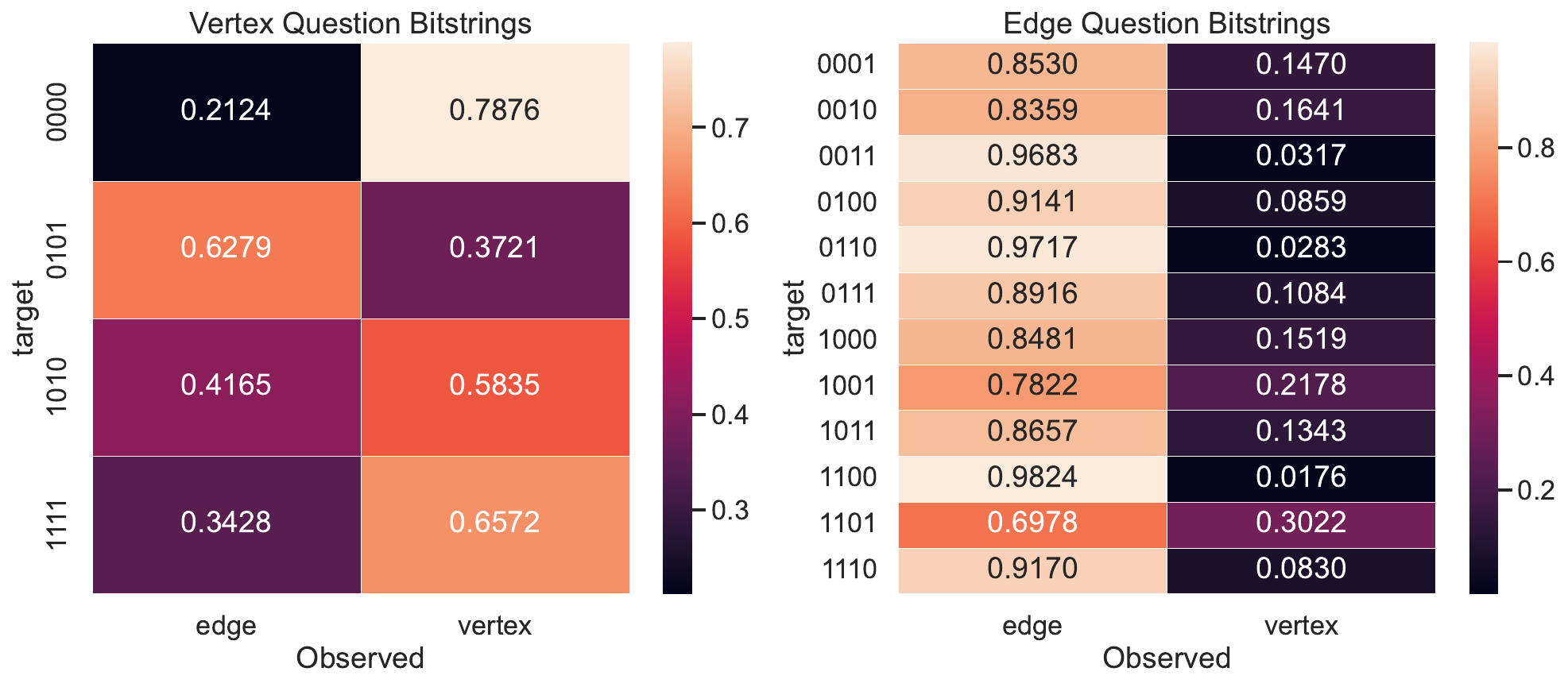}
        \caption{Rigetti \texttt{Ankaa-2}}
        \label{fig:Zbasis_SPAM_ankaa2}
        \end{subfigure}
    \centering
    \begin{subfigure}{\linewidth}
        \centering
        \includegraphics[width=\linewidth]{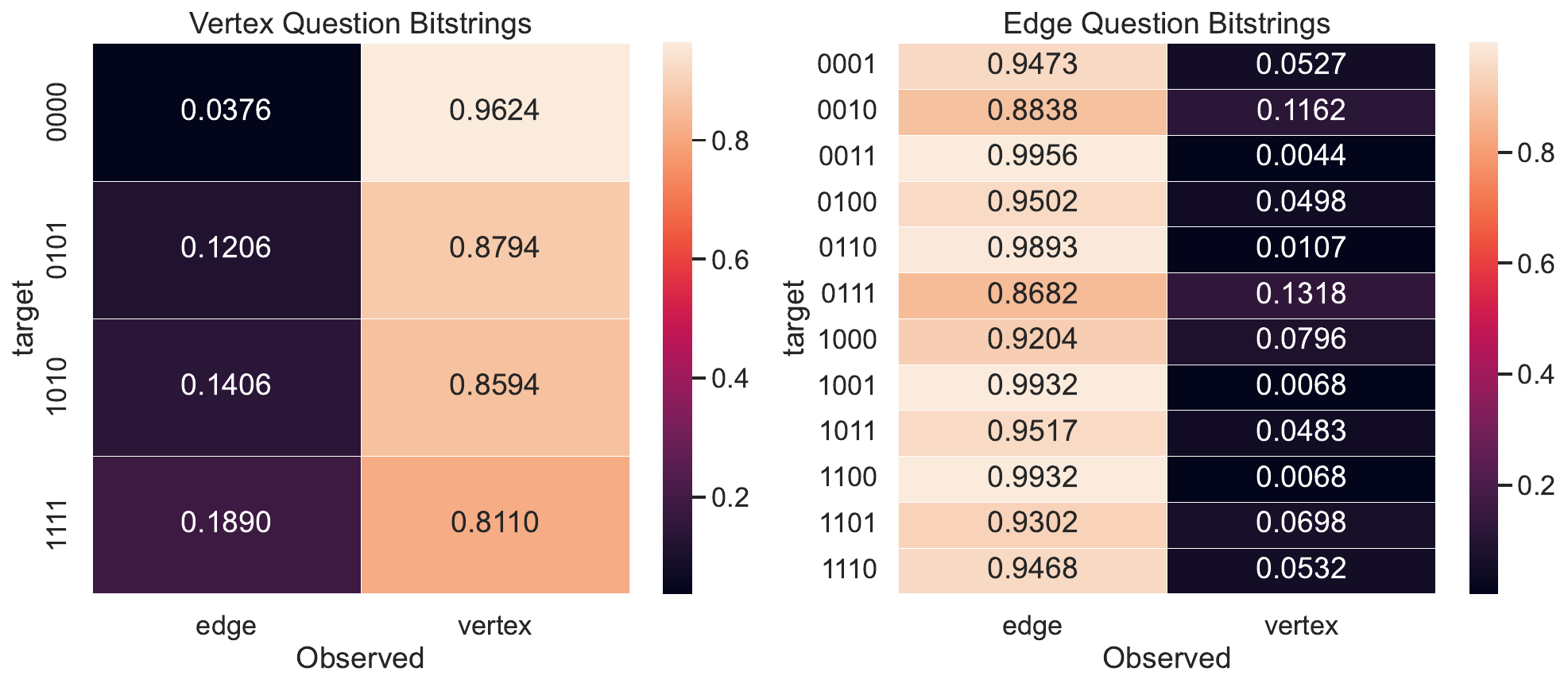}
        \caption{Rigetti \texttt{Ankaa-3}}
        \label{fig:Zbasis_SPAM_ankaa3}
        \end{subfigure}
    \centering
    \begin{subfigure}{\linewidth}
        \centering
        \includegraphics[width=\linewidth]{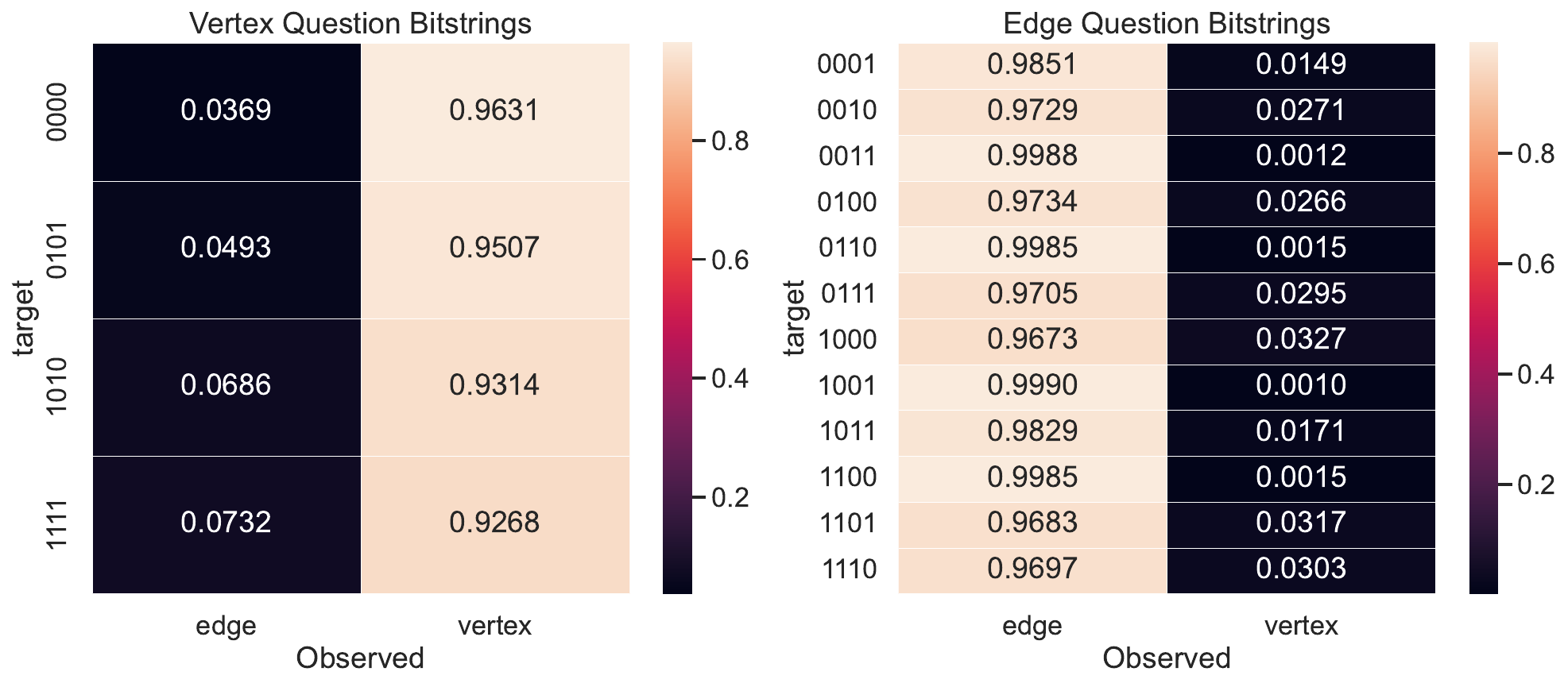}
        \caption{\texttt{ibm\_sherbrooke}}
        \label{fig:Zbasis_SPAM_sherbrooke}
        \end{subfigure}
\caption{Example of spurious bitstring counts caused by SPAM errors.}
\label{fig:spam_evd_ankaa}
\end{figure}

In this section we analyze how hardware noise affects different nonlocal game circuit components, supported by results collected on superconducting qubit platforms.  This extends the simulated noise results shown in Fig. \ref{fig:g14_noise_simulation} where the error rate of two-qubit gates was inferred from the hardware results reported in Section \ref{sec:hw_results}. 
We supplement these results with additional experiments designed to characterize key components of the strategy: readout measurement error, independent entangled measurements, and imperfect resource state preparation (shown in Fig. \ref{fig:noise_char}).  Throughout this section we analyze and characterize each element individually. We determine the effective win rate that would be observed by the players if one of these elements failed or was replaced by randomness and use this to demonstrate the effectiveness of nonlocal games as a hardware benchmark.
The readout measurement error can be characterized by a $2^n \times 2^n$ dimensional matrix constructed row-wise from individual computational basis state preparation and measurement: preparing the register in $|0\rangle^{\otimes n}$, applying $X$-gates, and projecting the final state onto the computational basis.  This can be used to estimate bit flip error probabilities (independent or correlated) \cite{hamilton2020scalable}, and also can be leveraged for readout error mitigation \footnote{The results we report do not include readout error mitigation, we reserve this for future work.}. We collected data to characterize the readout error on \texttt{ibm\_sherbrooke} and Rigetti's \texttt{Ankaa-2}, and \texttt{Ankaa-3} platforms. In Fig. \ref{fig:spam_evd_ankaa} we plot the \texttt{Ankaa-2} and \texttt{Ankaa-3} results to emphasize the connection to the EHD metric (see Section \ref{sec:hamming_bound}). Though the circuits executed on the hardware are very shallow, SPAM error can have a significant impact.

Connecting the SPAM error back to the EHD if a nonlocal circuit was correctly executed, and the only errors occurred during the readout stage, in Fig. \ref{fig:spam_evd_ankaa} we see that vertex questions are more likely to return incorrect answers, while for edge questions, correct answers can still be returned with high probability.  For vertex questions, the all-zero bitstring is relatively unaffected by errors during the readout measurement step, in contrast to the remaining three bitstrings.  For edge questions, bit-flip errors that occur during the readout step can still return valid edge question bitstrings.  The edge question in which the probability of erroneously returning a non-valid bitstring are bitstrings with high Hamming weights.  Thus, if a state is correctly prepared and the error only occurs during the readout stage, it affects vertex questions and low-weight edge answers.
The full quantum strategy is composed of multiple circuits needed to evaluate the players' performance on all questions posed by the referee. The construction assumes that the two players are separated in space to prohibit classical communication, and implementing the strategy requires nonlocal operations. Prior to the final qubit readout, the two players implement entangled unitaries ($\mathcal{U}_A \otimes \mathcal{U}_B$) that are assumed to be independent. We assess the ability of each player to apply these entangled measurements with high fidelity independently, and simultaneously without corrupting each others operations.  This is tested on four qubits $[46, 47, 48, 49]$ connected in a linear chain (see \ref{app: experimental_details}). A specific Bell state is prepared by applying  $\mathcal{U}_A \otimes 1 \otimes 1 $ or $1 \otimes 1 \otimes \mathcal{U}_B$ where only two qubits prepare a Bell state while the other two qubits remain in the $|0\rangle$ state.  Then, a Bell basis measurement is applied to the prepared Bell state and the remaining two qubits are measured in the computational basis.  This is compared to the preparation of two independent Bell states both measured by Bell state measurement. To amplify the gate noise we construct and execute these circuits with basic unitary folding by inserting pairs of CNOT gates. 

\begin{figure}
    \centering
     \centering
    \begin{subfigure}[b]{0.48\linewidth}
        \centering
        \includegraphics[width=\linewidth]{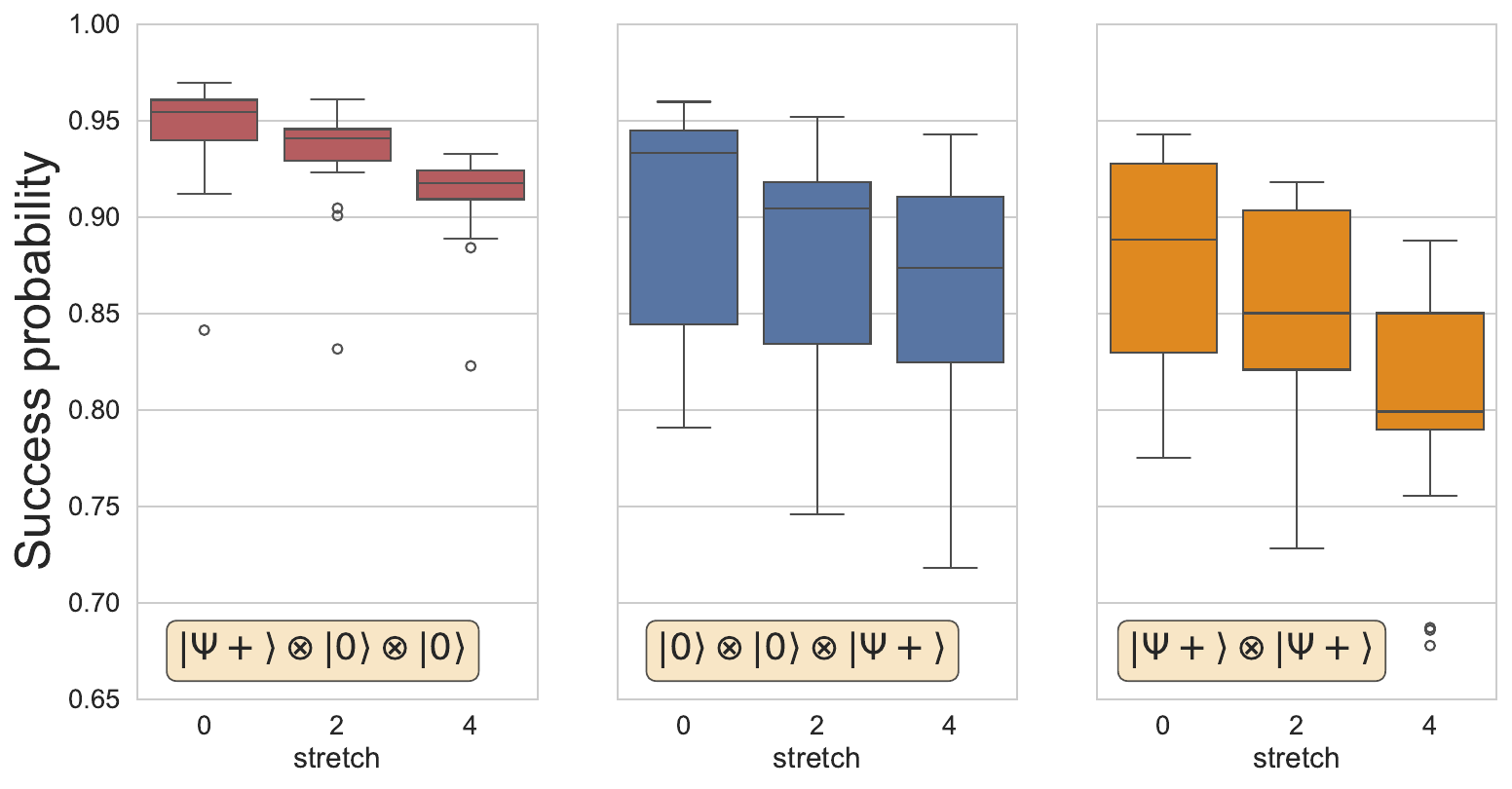}
        \caption{ $|\Psi+\rangle$ (\texttt{ibm\_sherbrooke}).}
        \label{fig:Psi_sherbrooke}
        \end{subfigure}
    \centering
    \begin{subfigure}{0.48\linewidth}
        \centering
        \includegraphics[width=\linewidth]{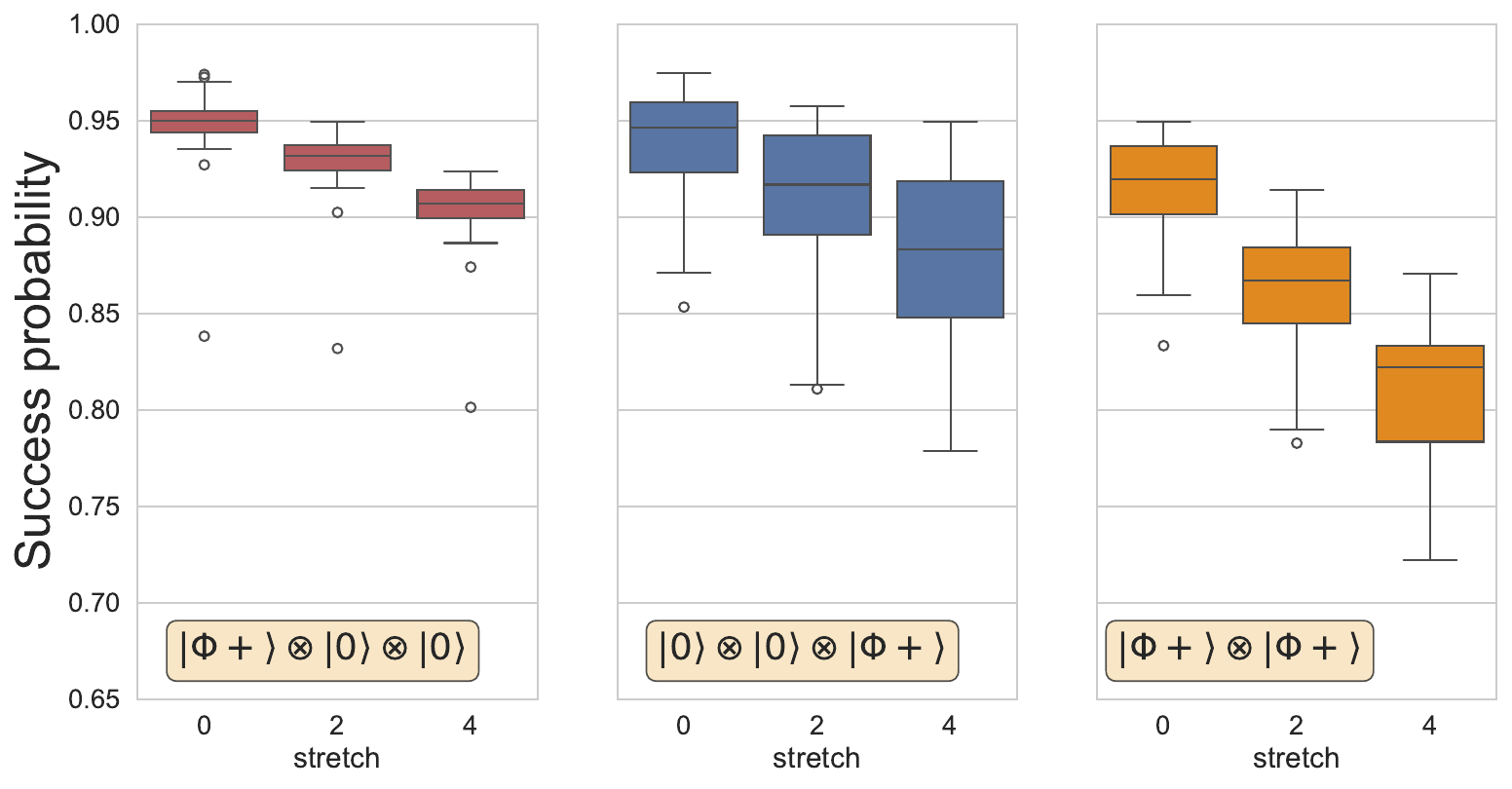}
        \caption{$|\Phi+\rangle$ (\texttt{ibm\_sherbrooke}).}
        \label{fig:Phi_sherbrooke}
        \end{subfigure}
\caption{Median probability of successfully preparing and measuring independent copies of Bell states ($|\Psi+\rangle$) or ($|\Phi+\rangle$) on \texttt{ibm\_sherbrooke}.}
\label{fig:independent_unitary_bell_states_sherbrooke}
\end{figure}

The general success probabilities are plotted in Fig. \ref{fig:independent_unitary_bell_states_sherbrooke}. 
For the single Bell state preparations, we extract the marginal distributions of each subset and plot the mean probability of observing counts of each Bell state.  The mean is evaluated using 14 executions of these experiments on  \texttt{ibm\_sherbrooke}.
\begin{figure}
    \centering
     \centering
    \begin{subfigure}[b]{\linewidth}
        \centering
        \includegraphics[width=\linewidth]{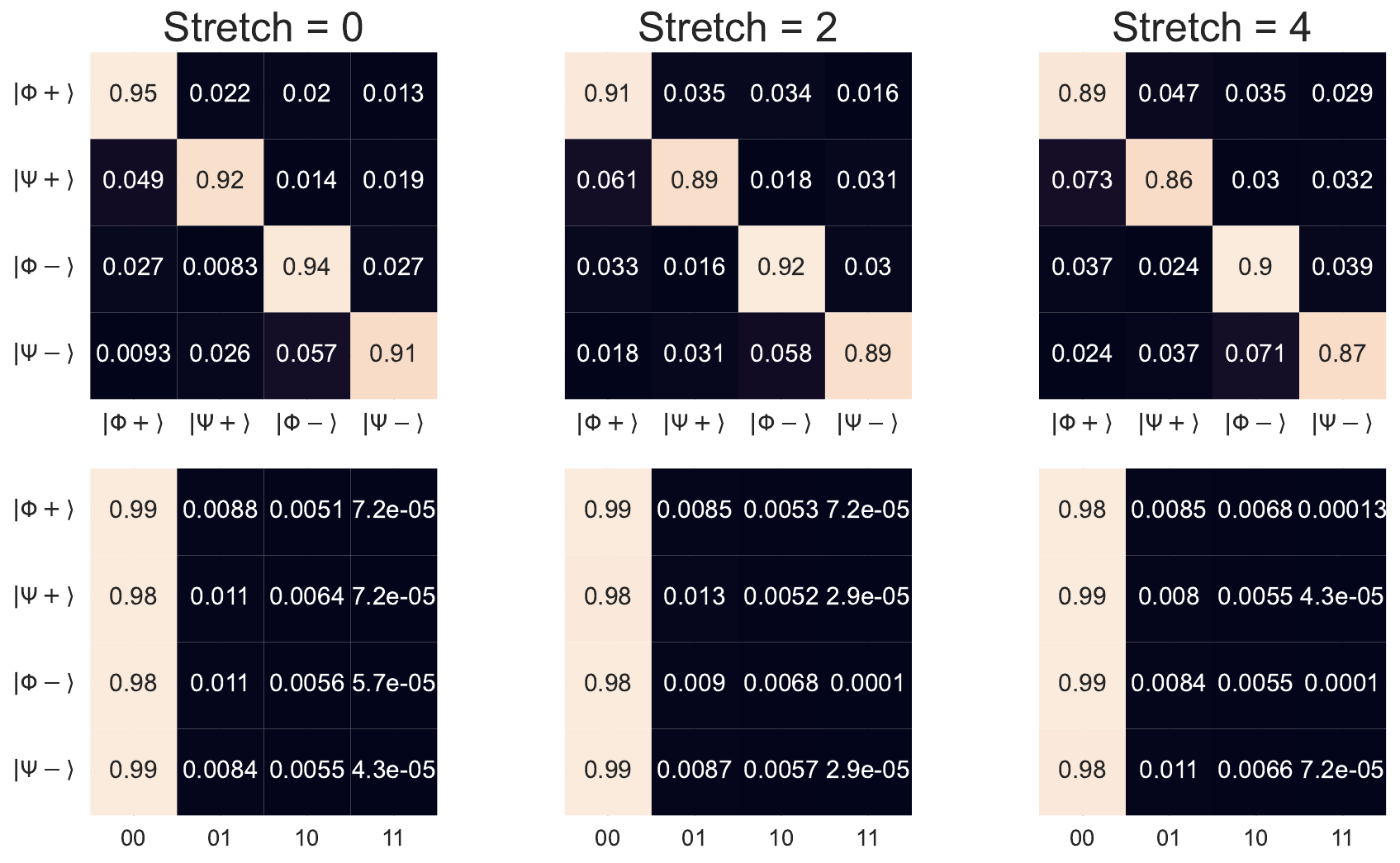}
        \caption{(Top) Bell states prepared on qubit subset $\mathcal{A}$. (Bottom) Qubit subset $\mathcal{B}$ remains idle.}
        \label{fig:UA_active_sherbrooke}
        \end{subfigure}
    \centering
    \begin{subfigure}[b]{\linewidth}
        \centering
        \includegraphics[width=\linewidth]{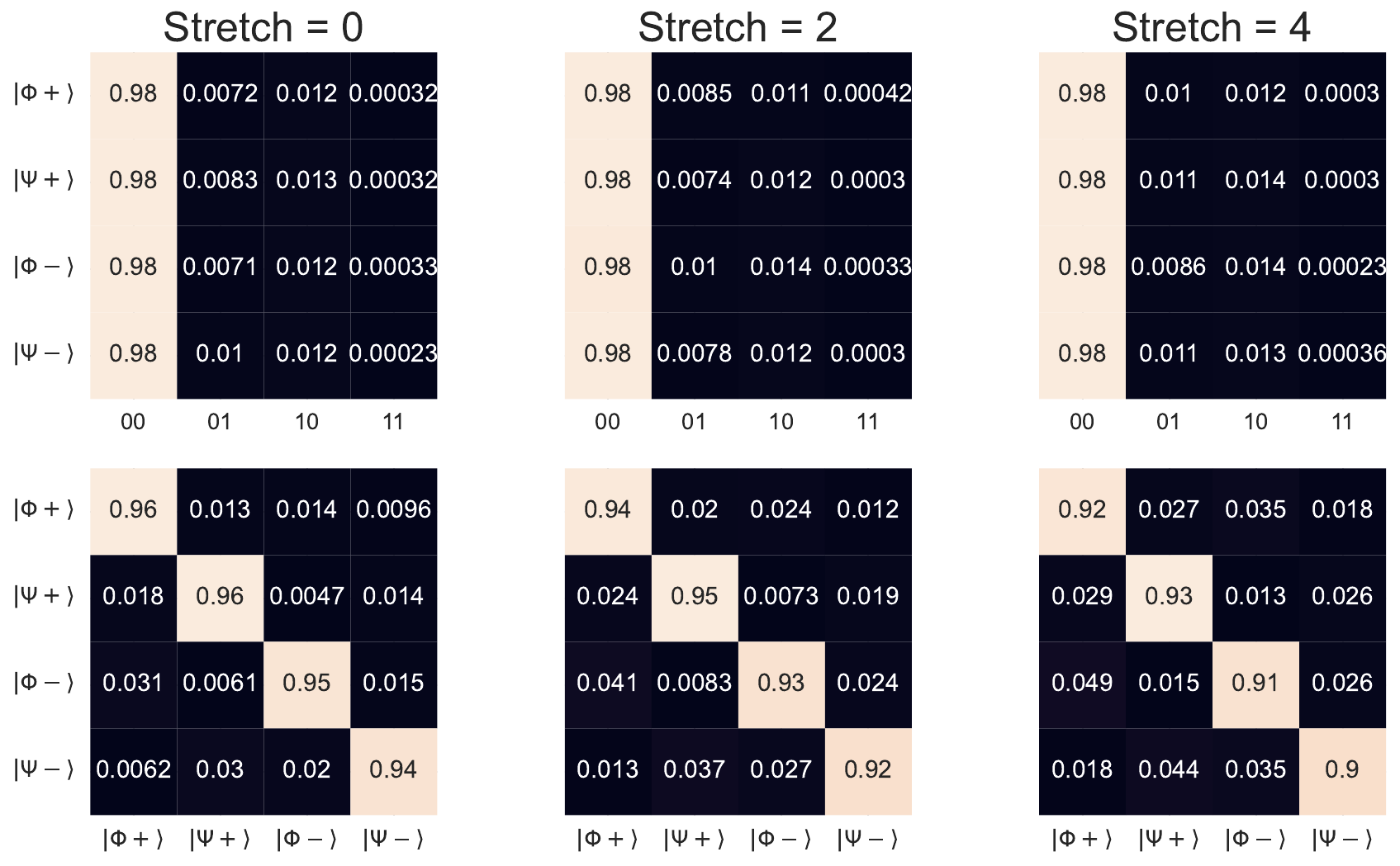}
        \caption{(Top) Qubit subset $\mathcal{A}$ remains idle. (Bottom) Bell states prepared on qubit subset $\mathcal{B}$.}
        \label{fig:UB_active_sherbrooke}
        \end{subfigure}
\caption{Mean probability of successfully preparing independent copies of Bell states combined with Bell state measurements. Mean probability of successfully observing idle qubits in the $|00\rangle$ state.}
\label{fig:sherbrooke_subsets}
\end{figure}
The distinct separation between the success probabilities of isolated Bell state preparation either on $[q_a,q_b]$ or $[q_c,q_d]$ could be caused by individual two qubit gate error rates -- indicative that a coupler between particular pairs of qubits could be more stable compared to neighboring qubits.  Another cause could be the choice of hardware qubits combined with circuit optimization options (see \ref{app: experimental_details}).

On \texttt{Ankaa-2} we prepared the state $|\Psi^+\rangle \otimes |0\rangle \otimes |0\rangle$ and observe that over 75\% of the observed bitstrings correspond to the correct Bell state.  The highest number of counts are returned in the all-zero bitstring, indicated that the state was prepared correctly and measured correctly while the idle qubits remained idle.  Preparing the state $|0\rangle \otimes |0\rangle \otimes |\Psi^+\rangle$ we observe that between 71-72\% of the observed bitstrings correspond to the correct Bell state.  However, preparing and measuring the state $|\Psi^+\rangle \otimes |\Psi^+\rangle$ showed a sharp decline in counts observed in the expected bitstrings. 
\begin{table}
    \centering
    \begin{tabular}{ccccc}
    \hline
      Stretch   & $|\Phi+\rangle_A$ & $|\Phi+\rangle_B$ & $|\Psi+\rangle_A$  & $|\Psi+\rangle_B$\\
       0  & 0.94 (1) & 0.96(2) & 0.91(2) & 0.96(3)\\
       2  & 0.92 (1) & 0.94(2) & 0.89(2) & 0.94(3)\\
       4  & 0.89 (2) & 0.90(2) & 0.86(2) & 0.92(4)\\
    \end{tabular}
    \caption{Mean and standard error of measuring counts in the target Bell state.}
    \label{tab:simultaneous_Bell_SPAM}
\end{table}
Connecting this characterization back to the nonlocal game as a benchmark: the game construction assumes the players are separated in space and classical communication is not possible. However the implementation on near-term hardware will likely use physical qubits that are connected via tunable couplings.  If correlated noise is significant when executing simultaneous multi-qubit gates on non-overlapping qubit subsets, this can affect the win rate of the players.  For the Bell state example we observe that this affects the ability to implement and measure two identical states. 
 We believe that correlated noise may impede the performance again of vertex questions.
Finally, we consider the impact of hardware noise on the resource state $|\Psi\rangle$ shared by Alice and Bob.  With mirrored unitary circuits \cite{mayer2021theory}, we measure the probability of applying $\mathcal{U}_{R}\mathcal{U}^{\dagger}_{R}$ and successfully returning to the initial all zero register.  Testing the four qubit unitary on \texttt{ibm\_sherbrooke}, \texttt{Ankaa-2}, and \texttt{Ankaa-3} multiple times we find that the success rate fluctuates depending on: hardware, qubit subset, and the choice of resource state.

On \texttt{ibm\_sherbrooke} the success rate of the mirrored four qubit unitary was 19.43\%. On \texttt{Ankaa-2}, the mirrored four qubit unitary of the original $G_{14}$ strategy, this approach had a success rate $<10\%$.  Specifically on September 29, 2024 the mirrored unitaries successfully returned to the initial state $|0\rangle^{\otimes 4}$: on qubit subset (9,10,17,16) 8.06\%; (2,3,10,9) 6.49 \%;  (9,10,16,17) 5.66 \%; (2,9,16,23) 8.06\%; and (2,3,9,10) 7.32 \%.  The circuit on \texttt{Ankaa-2} were compiled with PARTIAL re-wiring. For \texttt{Ankaa-3}, the mirrored four qubit unitary success rate was much higher.  On September 30, 2024 the mirrored unitaries successfully returned to the initial state $|0\rangle^{\otimes 4}$: (0,1,4,3) 55.32\%; (0,1,3,4) 33.08\%.  The circuits on \texttt{Ankaa-3} were compiled with NAIVE re-wiring. 

The mirrored circuits are much deeper than the resource state preparation alone, and contain more multi-qubit operations. Since noisy hardware can better prepare shallower, sparser resource state constructions, the mirror fidelity provides a pessimistic estimate of the fidelity of the resource state preparation.  However we find it informative to compare the mirror fidelity of the arbitrary four qubit unitary to the mirror fidelity of the shared states used in the Bell state strategy, which we measured $14$ times during one week using \texttt{ibm\_sherbrooke}.  For this set of shared states the mean success probability was $91.56\pm 0.91\%$. 

Overall what we can infer from these individual characterizations is how hardware can generate nonlocal correlations (in the resource state preparation), how independent qubits can be controlled (via the players entangled operations) and finally the robustness of the players answers to readout errors. The development of a full predictive model is beyond the scope of this work, but from the initial characterizations of the game components it is clear that improving individual components can significantly impact the overall win rate which is of importance in the $G_{14}$ game, where the separation between the classical and quantum strategies is small.

\section{Statistical fluctuations and Sample Complexity of Estimating the Win Rate}
\label{sec:sample_complexity}
On near-term quantum devices, the win rate of each circuit (question) is estimated by statistical sampling, using independently drawn samples to estimate the probability that the players return the correct answers. Finite sample effects lead to statistical fluctuations. In this section, we derive an upper bound on the number of individual samples (shots) to draw from a prepared state to sufficiently assess whether a circuit has correctly answered the referee's question.  

In the interactive nonlocal game setup, the scenario is repeated with random questions until the referee is satisfied with the outcome. We consider how to obtain a low error estimate of the win rate with high probability using a finite number of repetitions.

Let $n$ be the number of \textit{rounds} performed, where each round consists of the referee asking the players all $m$ possible questions once and checking their answers using the rule function $\lambda(a|q)$. In the context of quantum hardware, this can be viewed as the execution of $m$ quantum circuits with $n$ shots per circuit.

Because the outcome of each question is binary, i.e., $\lambda(a|q)\in \{0, 1\}$, we model the outcome of question $q_j$ as a Bernoulli random variable $\lambda_j$ with an unknown success probability $p_j$. The random variable describing the game value of a single round is $\omega = \frac{1}{m} \sum_j^m \lambda_j$. We denote the empirical estimate of the win rate with $n$ rounds as $\bar{\omega} = \frac{1}{n} \sum_i^n \omega_i$ where $\omega_1, \dots, \omega_n \sim \omega$ are i.i.d. samples. Under these mild assumptions, we derive an expression for the number of samples needed to accurately estimate the win rate within error $\epsilon$.

\begin{theorem}
    Let $\bar{\omega} = \frac{1}{n} \sum_i^n \omega_i$ be the empirical estimate of the game win rate after $n$ rounds, where each round $\omega_i$ is independent and identically distributed (i.i.d.). Then, for any $\epsilon > 0$,
    \begin{equation}
        P(|\bar{\omega} - \mathbb{E}[\bar{\omega}]| \geq \epsilon) \leq 2\exp\left(\frac{-n\epsilon^2/2}{\bar{\sigma}^2/m + \epsilon / 3}\right),
    \label{eq:win_rate_tail_prob}
    \end{equation}
    where $m = |Q|$ is the number of questions and $\bar{\sigma}^2 = \frac{1}{m}\sum_j^m p_j(1-p_j)$, where $p_j$ is the win rate of question $q_j$.
\end{theorem}

\begin{proof}
    We make use of the Bernstein inequality \cite{bernstein1924modification, zhang2021concentration}, which is restated here for convenience. Let $S_n = \sum_i^n X_i$ be the sum of zero-mean random variables $X_1, \dots, X_n$ and $|X_i| \leq c$ almost surely. Then, for any $\epsilon > 0$,

    \begin{equation}
        P(|S_n| \geq \epsilon) \leq 2\exp\left(\frac{-\epsilon^2/2}{\sum_i^n \mathrm{Var}[S_i] + c\epsilon/3}\right).
    \end{equation}
    
    To use the inequality, we construct the sum $S_n = \sum_i^n \omega_i - \mathbb{E}[\omega_i]$, subtracting the expectation values to meet the zero-mean condition, yielding

    \begin{equation}
        P(|n\bar{\omega} - n\mathbb{E}[\bar{\omega}]| \geq \epsilon) \leq 2\exp\left(\frac{-\epsilon^2/2}{\sum_i^n \mathrm{Var}[\omega_i] + c\epsilon/3}\right).
    \end{equation}

     The magnitude of each term is bounded $|\omega_i - \mathbb{E}[\omega_i]| \leq 1 = c$. Furthermore, because each round $\omega_i \sim \omega$ is i.i.d., $\sum_i^n \mathrm{Var}[\omega_i] = n\bar{\sigma}^2/m$, where $\bar{\sigma}$ is defined above and we have used the fact that the variance of a Bernoulli random variable is $p(1-p)$. Substituting $n\epsilon$ in place of $\epsilon$ gives (\ref{eq:win_rate_tail_prob}).
\end{proof}

\begin{corollary}[Sample complexity] With probability $1 - \delta$, we obtain an $\epsilon$-close estimate of $\omega$ using at least
    \begin{equation}
        n \geq 2\log(2/\delta)\left(\frac{\bar{\sigma}^2}{m\epsilon^2} + \frac{1}{3\epsilon}\right)
    \label{eq:win_rate_sample_complexity}
    \end{equation}
    rounds.

    \begin{proof}
        This results from setting (\ref{eq:win_rate_tail_prob}) less than or equal to $\delta$ and solving for $n$.
    \end{proof}
\end{corollary}

\begin{corollary}\label{confidence}
    [Confidence interval] With $n$ rounds and with probability $1 - \delta$, the error of our estimate is
    \begin{equation}
        |\bar{\omega} - \mathbb{E}[\bar{\omega}]| \leq \frac{2\log(2/\delta)}{3n} + \bar{\sigma}\sqrt{\frac{2\log(2/\delta)}{mn}}.
        \label{eq:win_rate_sample_error}
    \end{equation}

    \begin{proof}
        This can be obtained by solving (\ref{eq:win_rate_sample_complexity}) for $\epsilon$ and taking the positive solution, then applying the identity $\sqrt{x+y} \leq \sqrt{x} + \sqrt{y}$.
    \end{proof}
\end{corollary}

\begin{table}[htb!]
    \centering
    \input{overall_results_table}
    \caption{Overall win rate for each device with $95\%$ confidence intervals.}
    \label{tab:overall_results}
\end{table}

 From (\ref{eq:win_rate_sample_complexity}), we see two possibilities to achieve asymptotic $O(\log(1/\delta)/\epsilon)$ sampling: with a large number of questions $m$ and when $\bar{\sigma} \simeq \epsilon$. The first case is not practical because increasing the number of questions counterproductively increases the total number of circuit samples $mn$. 

The second case is also hard to achieve (at present) because it requires a near-perfect strategy on high-fidelity quantum hardware. This results from $\bar{\sigma}$ being directly linked to the win rate of the questions, which depends on both the strategy and quantum hardware. Assuming all questions have equal win rate $p$ for simplicity, this requires (again taking the positive solution) $p \approx \frac{1}{2}(1 + \sqrt{1 - 4\epsilon^2})$, which is approximately $p \approx 1 - 2\epsilon + O(\epsilon^2)$. We expect that $O(\log(1/\delta)/\epsilon)$ sampling may become feasible for perfect strategies with improved gate fidelity, quantum error correction, or amplitude amplification. Table \ref{tab:overall_results} contains all the win rates of our executed experiments with confidence intervals derived from Corollary \ref{confidence}.

\section{Conclusion}
\begin{figure}[!htb]
    \centering
    \includegraphics[width=\linewidth]{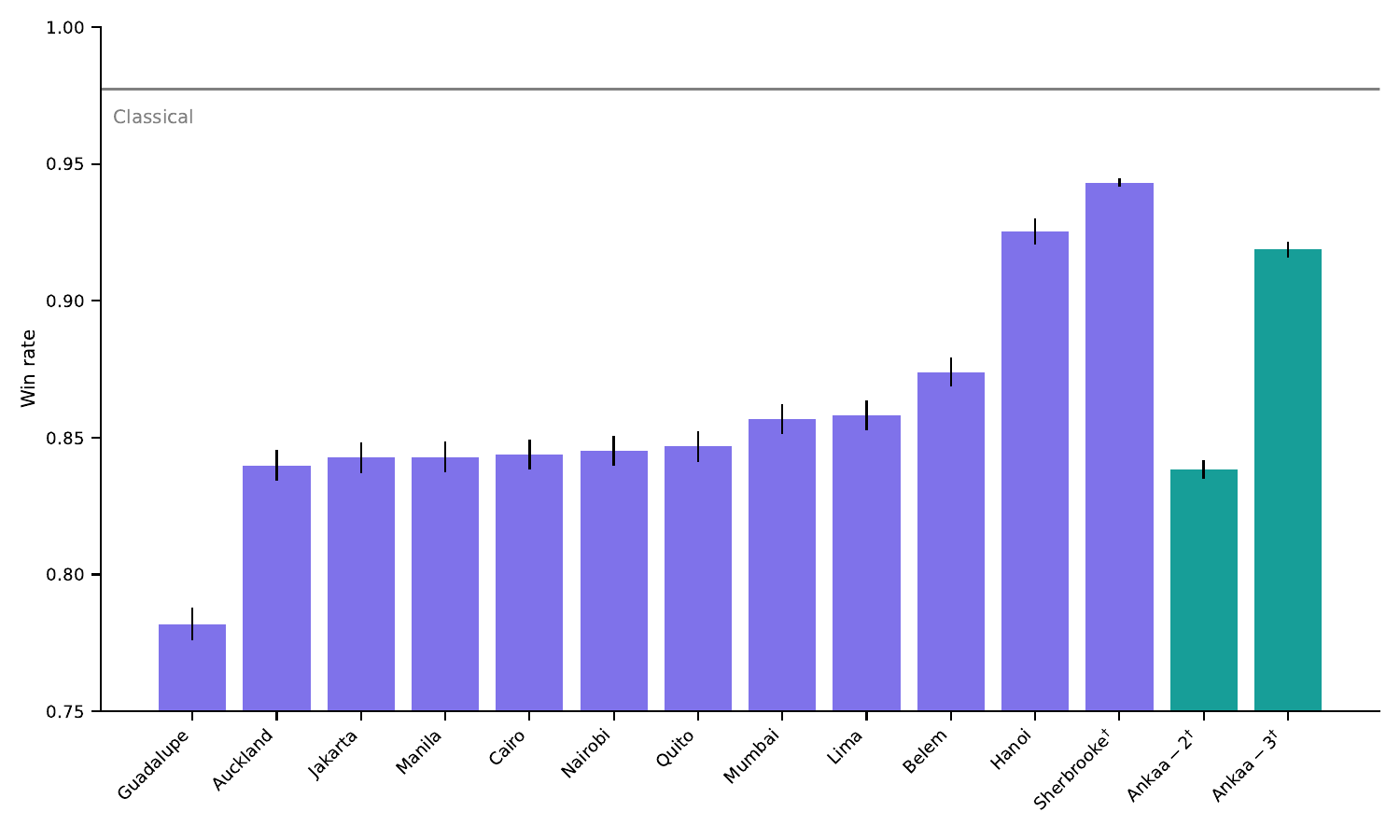}
    \caption{Overall win rate for all devices tested. The symbol $\dagger$ means that trial used the Bell pair strategy instead of the original strategy. Uncertainty on the win rates are 95\% confidence intervals. Colors represent the two device providers we tested: IBM in purple and Rigetti in teal. The horizontal line represent classical winrate threshold for this game; any strategy generated above this line requires quantum resources. All results are above the win rate achievable by random guessing.}
    \label{fig:overall_benchmark_results}
\end{figure}

We present a variational algorithm to compute novel quantum strategies for nonlocal games by encoding the rules of a nonlocal game into a Hamiltonian and employing a two-step optimization procedure. Our key insight is to optimize separately the state preparation circuit and the measurement scheme while leveraging robust circuit initialization and general techniques, such as ADAPT, during optimization. The proposed algorithm successfully reproduces known quantum strategies and has also discovered new short-depth, perfect quantum strategies for a graph on $14$ vertices using four qubits. This demonstrates that variational techniques can be effectively used on classical computers to identify short-depth, optimal strategies for small examples of nonlocal games where analytic methods fail. Moreover, these techniques extend to a quantum setting, where sample-based gradient estimation is employed. However, the presence of barren plateaus is a known challenge with the training objective function, suggesting that ``warm starts" or other techniques to mitigate vanishing gradients may be necessary for scaling these methods to larger nonlocal games.

We further illustrate how the execution of a nonlocal game strategy can serve as an application-level benchmark for quantum devices. By evaluating the win rates of both vertex and edge questions in these games, the win rate of vertex questions reflects a device's ability to perform nonlocal operations and maintain gate fidelity, while the win rate of edge questions can help confirm the utilization of entanglement across a device. Although none of the devices we tested surpassed the quantum advantage threshold, primarily due to noise in circuit execution, we believe our results can be improved by optimizing the transpilation of the individual circuit before execution and control of the device calibration schedules. It is also worth noting that although our experiments do not provide a full proof of quantum advantage, given that the particles are not spatially separated enough to guarantee that classical communication does not happen during the experiment, it does provide validation that the quantum hardware in question outputs results consistent with the hypotheses of quantum theory. Recent work has begun to outline ways of guaranteeing a ``loop-hole free'' full verification of quantum advantage by compiling a multi-prover nonlocal game strategy into a single prover strategy \cite{grilo2019simple, natarajan2023bounding, kalai2023quantum} and we leave it to future work to investigate the feasibility and implications of these schemes for the games we studied. In a recent survey \cite{acuaviva2024benchmarking}, the authors outlined five desirable properties for a good quantum benchmark and in our work we argued how the win rate from nonlocal game strategies fit all five points:
\begin{itemize}
    \item {\bf Relevant:} The win rate measures the ability to prepare, control, and manipulate entangled states.
    \item {\bf Reproducible:}  Strategy and questions are fixed.
    \item {\bf Fair:} Device independent and the executed circuits are shallow. 
    \item {\bf Verifiable:} Straightforward to calculate the win rate via sampling.
    \item {\bf Usable:} Circuits can be made accessible via QASM files and can easily be ported to other quantum devices.
\end{itemize}
We believe that the continued study and extensions of nonlocal games, in particular graph-based games, can enable the design of more appropriate quantum benchmarks as quantum devices scale and hardware architectures become more complex. Ultimately, our research not only advances the understanding of variational quantum strategies but also lays the foundation for leveraging quantum machine learning techniques to explore other nonlocal games strategies beyond the reach of classical methods.

\section*{Acknowledgments}
Thanks to David Roberson and Eleanor Rieffel for providing valuable feedback. NW, JF, and COM were funded by grants from the US Department of Energy, Office of Science, National Quantum Information
Science Research Centers, Co-Design Center for Quantum Advantage under contract number DE-SC0012704. JF and COM were partially supported by the Laboratory Directed Research and Development Program and Mathematics for Artificial Reasoning for Scientific Discovery investment at the Pacific Northwest National Laboratory, a multiprogram national laboratory operated by Battelle for the U.S. Department of Energy under Contract DEAC05- 76RLO1830. S. C. is supported in part by the DOE Advanced Scientific Computing Research (ASCR) Accelerated Research in Quantum Computing (ARQC) Program under field work proposal ERKJ354. K. H. was supported by the DOE Advanced Scientific Computing Research (ASCR) Pathfinder Testbed Program under FWP ERKJ418.

This research used resources of the Oak Ridge Leadership Computing Facility, which is a DOE Office of Science User Facility supported under Contract DE-AC05-00OR22725.

This manuscript has been authored in part by UT-Battelle, LLC, under Contract No. DE-AC0500OR22725 with the U.S. Department of Energy. The United States Government retains and the publisher, by accepting the article for publication, acknowledges that the United States Government retains a non-exclusive, paid-up, irrevocable, world-wide license to publish or reproduce the published form of this manuscript, or allow others to do so, for the United States Government purposes. The Department of Energy will provide public access to these results of federally sponsored research in accordance with the DOE Public Access Plan.

\section*{References}
\bibliographystyle{vancouver}	
\bibliography{qref}
\newpage

\appendix
\section*{Appendix}

\section{Data Availability}\label{app: data}

The code used to generate the data and figures in this article can be found at \\ \href{https://github.com/jfurches/nonlocalgames}{\texttt{https://github.com/jfurches/nonlocalgames}}.  The authors will make available the data collected for noise characterization by reasonable request. 

\section{ADAPT-VQE} \label{app: vqe}

The Adaptive Derivative-Assembled Pseudo-Trotter ansatz Variational Quantum Eigensolver (ADAPT-VQE) is a hybrid quantum-classical algorithm designed to dynamically construct an efficient and compact ansatz for molecular simulations on quantum hardware \cite{grimsley2019adaptive}. It enhances the traditional Variational Quantum Eigensolver (VQE) by adaptively building a problem-specific ansatz for the quantum state. Unlike traditional approaches such as Unitary Coupled Cluster (UCC), which rely on pre-defined and often redundant wavefunction ansätze, ADAPT-VQE grows the ansatz iteratively by selecting operators that maximize energy recovery at each step. This adaptive approach minimizes the number of parameters and quantum gates required, making it well-suited for noisy intermediate-scale quantum (NISQ) devices.

ADAPT-VQE operates by measuring the gradient of the Hamiltonian's expectation value with respect to each operator in a predefined operator pool. The operator with the largest gradient is added to the ansatz, and its parameter is optimized alongside previously added parameters using a classical variational optimizer. This process is repeated until the norm of the gradient vector falls below a threshold, ensuring convergence to the desired accuracy. 

More concretely: assume we have variational parameters $\vect{\theta}^{(k)} = (\theta_1,\dots,\theta_k)$ and the operator pool $\mathcal{A} = \{ A^{(1)}, A^{(2)}, \dots A^{(N)}\}$, the ansatz in iteration $k+1$ of the algorithm may be written as
\begin{equation*}
    |\psi_{k+1}(\vect{\theta}^{(k+1)})\rangle = e^{-i\theta_{k+1 }A_{k+1}}|\psi_{k}(\vect{\theta}^{(k)})\rangle.
\end{equation*}
Notice that the ansatz at iteration $k$ is grown by appending operator $A_{k+1}$ with coefficient $\theta_{k+1}$; the operator is chosen by measuring the energy gradients $\left| \left. \partial \langle H \rangle / \partial \theta_{k+1}\right|_{\theta_{k+1}=0} \right|$ for each operator in the pool and selecting the one with the largest gradient. For this step, it can be shown that 
\begin{equation*}
    \left| \left. \partial \langle H \rangle / \partial \theta_{k+1}\right|_{\theta_{k+1}=0} \right| =  \left| \langle \psi_{k}(\vect{\theta}^{(k)}) | \left[ A_{k+1}, H \right] | \psi_{k}(\vect{\theta}^{(k)}) \rangle \right|,
\end{equation*}
where the right hand side can be efficiently measured on a quantum processor as the size of a problem scales. The pool operator gradient-measurement step is followed by a convergence check: if the pool operator gradient norm is smaller than a threshold $\varepsilon$, the calculation is terminated; if not, the iteration procedure continues. The ansatz-growing step is followed by a VQE optimization of all variational parameters. 

By tailoring the ansatzs to the problem at hand, ADAPT-VQE achieves high accuracy with significantly reduced circuit depth compared to fixed ansatz methods. This variational technique has been studied extensively \cite{cerezo2021variational} and it has been extended to tackle problems in Quantum Generative training \cite{warren2022adaptive, sherbert2024adaptive} 

\section{Original $G_{14}$ Strategy}\label{app:strategy}

Here we outline the perfect quantum strategy for the graph $G_{14}$ using $4$ colors as detailed in \cite{MR2}. The authors construct this strategy by leveraging a 4-dimensional real orthogonal representation of the graph and a transformation derived from quaternion multiplication outlined in \cite{cameron2007quantum}. Here is an outline of their construction:
\begin{enumerate}
    \item {\bf Ortogonal Representation:} For each vertex in $v \in V(G_{14})$ we assign a normalized 4D real unit vector $\varphi(v)$ as follows:
    \begin{itemize}
        \item For each vertex in $G_{13}$ (see Figure \ref{fig:g13}) you assign it a 3-dimensional vectors with entries in $\{-1, 0, 1\}$ such that two vertices in $G_{13}$ are adjacent if and only if their corresponding vectors are orthogonal.
        \item For each 3-dimensional vector, $(x, y, z)^T$, extended it to a 4D vector by appending a zero: $(x, y, z, 0)^T$.
        \item Assign the apex vertex $\Omega$ of $G_{14}$ the 4D vector $(0, 0, 0, 1)^T$.
        \item Normalize each vector to be a unit vector and let $\varphi(v)$ be the vector corresponding to vertex $v\in G_{14}$. 
        \item This assignment guarantees that if vertices $u$ and $v$ are adjacent ($u \sim v$), their vectors are orthogonal ($\varphi(u)^T \varphi(v) = 0$). 
        \item To each vector $\varphi(v) = (r_0, r_1, r_2, r_3)^T$, we associate a set of four mutually orthogonal unit vectors, $\{\varphi_k(v)\}_{k=0}^3$, where each vector is a columns of the following matrix:
    $$
    M_v = 
    \begin{pmatrix}
    r_0 & -r_1 & -r_2 & -r_3 \\
    r_1 &  r_0 &  r_3 & -r_2 \\
    r_2 & -r_3 &  r_0 &  r_1 \\
    r_3 &  r_2 & -r_1 &  r_0
    \end{pmatrix}
    $$
    So, $\varphi_0(v) = (r_0, r_1, r_2, r_3)^T$, $\varphi_1(v) = (-r_1, r_0, r_3, -r_2)^T$, and so on. These four vectors form the measurement basis for vertex $v$.
    \end{itemize}

\item {\bf State and Projectors $P_k(v)$:} In the corresponding nonlocal game, Alice and Bob share a 4-dimensional maximally entangled state, $|\Psi^+\rangle = \frac{1}{2} \sum_{j=0}^{3} |j\rangle_A \otimes |j\rangle_B$. Upon receiving a vertex $v$, a player performs a measurement using projectors $\{P_k(v)\}_{k=0}^3$, where each projector is defined by the corresponding basis vector:
$$
P_k(v) = \varphi_k(v) \varphi_k(v)^T
$$

\item {\bf Joint Probabilities $P(a,b|u,v)$:} The probability that Alice and Bob obtain outcomes (colors) $a$ and $b$ for questions $u$ and $v$ respectively, is given by:
$$
P(a,b|u,v) = \frac{1}{4} \text{Tr}(P_a(u) P_b(v)) = \frac{1}{4} |\varphi_a(u)^T \varphi_b(v)|^2
$$
This formula ensures that the winning conditions of the coloring game are met with certainty. Specifically, if $u=v$, then $P(a,b|u,u) = \frac{1}{4}\delta_{ab}$. If $u \sim v$, then $P(a,a|u,v)=0$ for all $a$.
\end{enumerate}

One difficulty in implementing this strategy comes from the fact that measurement schemes are given as projections, which would need to be decomposed, for example, using Linear Combinations of Unitatires (LCU) \cite{childs2012hamiltonian}. The cost of standard LCU can be resource-intensive requiring $\left \lceil{\log(M)}\right \rceil$ ancilla qubits, where $M$ is the number of unitaries in the linear combination, as well as the need to implement the ``prepare'' unitary and a sophisticated multi-qubit controlled ``select'' unitary for each projection separately. Other techniques like Ancilla-free LCU \cite{chakraborty2024implementing} might be able to reduce this overhead, but assessing the feasibility of implementing this strategy in near-hardware is non-trivial and outside the scope of the work.

\section{Measurement Parameters}

Alice's measurement parameters of the $G_{14}$ strategy are contained within\\ 
\texttt{data/g14\_constrained\_u3ry/g14\_state.json} with the key \texttt{phi}. Constructing this into a NumPy array should return a tensor of shape $(1, 14, 2, 4)$, corresponding to \texttt{(players, questions, qubits, parameters)}. This tensor can be transformed to produce the conjugated measurement angles for Bob, as seen in \texttt{U3RyLayer} in \texttt{measurement.py}.

\section{Hyperparameters}

\begin{table}[h]
\centering
\begin{tabular}{c|lc}
\textbf{Problem} & \textbf{Hyperparameter} & \textbf{Value}     \\ \hline
\multirow{3}{*}{CHSH}     & ADAPT Grad Max $\epsilon_{\theta}$ & $10^{-3}$ \\
                          & BFGS Grad Max $\epsilon_{\phi}$    & $10^{-5}$ \\
                          & DPO Tolerance $\Delta E$           & $10^{-3}$ \\ \hline
NPS                       & Same as CHSH                       &           \\ \hline
\multirow{3}{*}{$G_{14}$} & ADAPT Grad Max $\epsilon_{\theta}$ & $10^{-6}$ \\
                          & BFGS Grad Max $\epsilon_{\phi}$    & $10^{-5}$ \\
                          & DPO Tolerance $\Delta E$           & $10^{-6}$\\
\end{tabular}
\caption{Hyperparameters for DPO experiments}
\label{tab:hyperparameters}
\end{table}

We give the algorithm hyperparameters for our experiments. The parameter $\epsilon_\theta$ refers to the convergence criteria of ADAPT used to prepare the shared state $\ket{\psi(\theta)}$. ADAPT finishes when the maximum pool gradient element reaches the threshold, $\max_{A_i} \left| \braket{[H,A_i]} \right| < \epsilon_\theta$. Similarly, the parameter $\epsilon_\phi$ controls the convergence of the second phase of DPO, as the BFGS optimizer halts when $\max_i \left| \nabla_\phi \braket{H}\right| < \epsilon_\phi$. Finally, $\Delta E$ controls the termination of the overall DPO procedure, ending when $\braket{H^{(k-1)}} - \braket{H^{(k)}} < \Delta E$ at iteration $k$.

\section{Gradient Sample Complexity}\label{app: grad_complexity}
In this section, we analyze the efficiency of the gradient simulation to understand its sample complexity. This addresses the practical and theoretical challenges faced when implementing our algorithm. The gradient complexity we consider is in terms of the number of exponentials required to achieve any $\epsilon$ precision. 
\begin{theorem}\label{thm:random}
    Let $\mathcal{E}_j$ be a random variable describing the error in the gradient estimate for the $j-th$ experiment with variance $\mathbb{E}[\mathcal{E}^2_j]=\epsilon_0^2$. Then the sample complexity of estimating the gradient with $\epsilon^2$ variance is given by $N_{\mathrm{exp}}\in\mathcal{O}\left(\frac{N^2}{\epsilon^2} \right)$, where $N$ is the dimensionality of the parameter space. 
\end{theorem}

\begin{proof}
    Let $\epsilon_0=\frac{\epsilon}{\sqrt{N}}$. By the additivity of the variance, it follows that $\mathbbm{E}\left[\sum\limits_{j=1}^N \mathcal{E}_j^2\right] = N\mathbbm{E}[\mathcal{E}_j^2]=N\epsilon_0^2$. The Euclidean norm of the gradient is approximated using the variances of the measurement outcomes. Hence 
    \begin{equation}
        \|\nabla\|^2\approx \mathbbm{E}\left[\sum\limits_{j=1}^N \mathcal{E}_j^2\right]=N\frac{\epsilon^2}{N}=\epsilon^2.
    \end{equation}
    Since each experiment requires $\mathcal{O}\left(\frac{1}{\epsilon_0^2}\right)=\mathcal{O}\left(\frac{N}{\epsilon^2}\right)$ operator exponentials and this must be repeated $N$ times, the total number of operator exponentials $N_\mathrm{exp}$ is 
    \begin{equation}
        N_\mathrm{exp}\in\mathcal{O}\left(\frac{N^2}{\epsilon^2}\right)
    \end{equation}
    as desired. 
\end{proof}

\begin{theorem}
    Assume that the variational state $\ket{\psi(\theta)}$ requires $N$ parameters to specify and that we wish to minimize $F(\theta):=\langle\psi(\theta)|H|\psi(\theta)\rangle$ over $\theta$.  Assume that $F$ is Lipshitz continuous with constant $C$ and that $\nabla F$ is Lipshitz continuous with constant $L$.  We then have that the number of exponentials required to perform gradient descent optimization with final error in the objective function at most $\epsilon_{\rm tot}$ using learning rate $\eta$ and $N_{\rm epochs}$ epochs is in
    $$
    O\left(\frac{N^2 N_{\rm epoch} C^2((1+\eta L)^{N_{\rm epoch}}-1)^2 }{\epsilon_{\rm tot}^2 L^2} \right)
    $$ 
    \begin{equation}
        N_{\mathrm{epoch, tot}}\in\mathcal{O}\left(\frac{N^2N_{\mathrm{epoch}}}{\gamma^2\min\limits_{\theta\in\Gamma}\|\nabla\langle\psi(\theta)|H(\phi)|\psi(\theta)\rangle \|}\right).
    \end{equation}
\end{theorem}

\begin{proof}
The gradient descent rule with learning rate $\eta$ reads
    \begin{equation}
        \theta \rightarrow \theta - \eta \nabla_{\phi}\langle\psi(\theta)|H(\phi)|\psi(\theta)\rangle.
    \end{equation}
Using our assumption that the gradient is Lipshitz-continuous with constant $L$, then
    \begin{equation}
        \|\nabla \langle\psi(\theta)|H|\psi(\theta)\rangle - \nabla\langle\psi(\theta+\delta)|H|\psi(\theta+\delta)\rangle\| \le L\delta.
    \end{equation}
If we define $\tilde{G}(\theta)$ to be an approximate  gradient evaluated at the parameters $\theta$, then
\begin{align}
    \|\nabla \langle\psi(\theta)|H|\psi(\theta)\rangle - \tilde{G}(\theta+\delta)\| \le& \|\nabla \langle\psi(\theta)|H|\psi(\theta)\rangle - \nabla\langle\psi(\theta+\delta)|H|\psi(\theta+\delta)\rangle\| \nonumber\\
    &\qquad+\| \nabla\langle\psi(\theta+\delta)|H|\psi(\theta+\delta) -\tilde{G}(\theta +\delta)\|\nonumber\\
    &\le L\|\delta\| + \epsilon.
\end{align}
Thus, we can recursively define the error in the parameter vector after $k$ epochs to be $\delta_k$ and thus from the triangle inequality and the gradient update rule we have
\begin{equation}
    \|\delta_k\| \le \eta(L\|\delta_{k-1}\| +\epsilon) + \|\delta_{k-1}\|
\end{equation}
We can then solve this recursion relation to find that
\begin{align}
    \|\delta_k\| &\le \eta \epsilon + (1+\eta L)\eta \epsilon + (1+\eta L)^2 \eta \epsilon + \cdots\nonumber\\
    &\le \frac{ \epsilon((1+\eta L)^k -1)}{L}.
\end{align}
Using the assumption that the objective function is Lipshitz-continuous with constant $C$,
\begin{align}
    \|\langle\psi(\theta+\delta)|H|\psi(\theta+\delta)\rangle -  \bra{\psi(\theta)} H \ket{\psi(\theta)}\le C \|\delta\|.
\end{align}
Then it suffices to choose the error per gradient evaluation such that
\begin{equation}
    \frac{C\epsilon (( 1+\eta L)^{N_{\rm epoch}} -1)}{L} \le \epsilon_{tot}.
\end{equation}
Isolating $\epsilon$ yields
\begin{equation}
    \epsilon \le \frac{\epsilon_{\rm tot} L }{C((1+\eta L)^{N_{\rm epoch}}-1)}.
\end{equation}
This means that from Theorem~\ref{thm:random}, the total number of exponentials per epoch that are needed is
\begin{equation}
    N_{\exp} \in O\left(\frac{N^2C^2((1+\eta L)^{N_{\rm epoch}}-1)^2 }{\epsilon_{\rm tot}^2 L^2} \right).
\end{equation}
Using the fact that there are $N_{\rm epoch}$ repetitions of the above 
\begin{equation}
    N_{\mathrm{\exp, tot}} \in O\left(\frac{N^2 N_{\rm epoch} C^2((1+\eta L)^{N_{\rm epoch}}-1)^2 }{\epsilon_{\rm tot}^2 L^2} \right).
\end{equation}

\end{proof}

This shows that the sample complexity of such problems can, in general, be substantial. In particular, if a small learning rate is required for the evolution, the number of operations needed for optimization can be exponential. The learning rate $\eta$ should be chosen (in the strongly convex case) to be proportional to the smallest eigenvalue of the Hessian matrix, implying that the number of samples scales exponentially with the condition number. This can be prohibitive in cases where some optimization directions are vastly steeper than others, such as in the vicinity of a saddle point. The number of epochs required for optimization is similarly difficult to bound. However, in the case where the optimization function is strongly convex, the number of epochs varies logarithmically with the error in the final objective function. In general, however, such optimization problems are not necessarily strongly convex. For these reasons, we leave the parameters of the gradient descent arbitrary.

As a final note, this suggests that variationally optimizing the parameters for a nonlocal game is not necessarily expected to be efficient, in general. To make this optimization tractable at scale, we need to minimize the number of epochs as much as possible. This can be achieved by starting with a well-informed initital guess for the protocol before attempting to optimize the result. If such conditions are met, the above analysis suggests that a manageable number of operations will be needed to achieve a constant distance from the locally optimized strategy. To tackle the general problem, we suggest exploring alternative optimization approaches such as solving the variational problem using dynamical simulation-based methods \cite{catli2025exponentially}. 

\section{Experimental Details}\label{app: experimental_details}
The experiments on \texttt{ibm\_sherbrooke} were conducted 7 different times between Sep. 27 - Oct. 1, 2024 with 4096 shots per circuit. The layout was chosen on the first run to be qubits 46-49 (a linear chain) using the \texttt{dense} method of the Qiskit transpiler with no optimization (level 0). For subsequent runs, the same layout was repeated. Each batch contained:  the SPAM characterization circuits, the independent unitary noise characterization circuits, mirror fidelity circuits and the Bell pair game circuits. In Table \ref{tab:overall_results} and Fig. \ref{fig:overall_benchmark_results}, the best run on \texttt{ibm\_sherbrooke} is reported. Calibration data for the backend was queried and saved at the time the circuit batch entered the queue and in Table \ref{tab:error_rates} we report the two qubit gate error (ECR gates).  

\begin{table}
    \centering
    \begin{tabular}{c|c}
       name  & value \\
    \hline
       ecr45\_44  & 0.010494 \\
        ecr45\_46 & 0.007731 \\
         ecr47\_46 & 0.004980 \\
         ecr47\_48 & 0.006505 \\
         ecr49\_48& 0.005589\\
         ecr49\_50 & 0.010321\\
         ecr50\_51 & 0.007020
    \end{tabular}
    \caption{Calibration data for individual ECR gates between hardware qubits 46-49 on \texttt{ibm\_sherbrooke}. }
    \label{tab:error_rates}
\end{table}

The reported data from Rigetti's \texttt{Ankaa-2} was collected on September 29 2024, and September 30 2024. Each circuit was sampled with 2048 shots and the hardware qubits used are reported in Figs. \ref{fig:ankaa2_G14} and \ref{fig:ankaa2_bell_pair}.  

The reported data from Rigetti's \texttt{Ankaa-3} was collected on September 30 2024
. Each circuit was sampled with 2048 shots and the hardware qubits used are reported in Figs. \ref{fig:ankaa3_G14} and \ref{fig:ankaa3_bell_pair}.

\end{document}

%% file: overall_results_table.tex
\begin{tabular}{lllccc}
\toprule
\bf{Year} & \bf{Provider} & \bf{Device} & \bf{Strategy} & \bf{Shots} & \bf{Win rate} (\%) \\
\midrule
\multirow[t]{11}{*}{2023} & \multirow[t]{11}{*}{IBM} & Guadalupe & Original & 1024 & 78.1(6) \\
 &  & Auckland & Original & 1024 & 83.9(6) \\
 &  & Jakarta & Original & 1024 & 84.2(6) \\
 &  & Manila & Original & 1024 & 84.2(6) \\
 &  & Cairo & Original & 1024 & 84.3(6) \\
 &  & Nairobi & Original & 1024 & 84.5(6) \\
 &  & Quito & Original & 1024 & 84.6(6) \\
 &  & Mumbai & Original & 1024 & 85.6(6) \\
 &  & Lima & Original & 1024 & 85.8(6) \\
 &  & Belem & Original & 1024 & 87.3(6) \\
 &  & Hanoi & Original & 1024 & 92.5(5) \\
\cline{1-6}
\multirow[t]{3}{*}{2024} & IBM & Sherbrooke & Bell Pair & 4096 & \textbf{94.3(2)} \\
\cline{2-6}
 & \multirow[t]{2}{*}{Rigetti} & Ankaa-2 & Bell Pair & 2048 & 83.8(4) \\
 &  & Ankaa-3 & Bell Pair & 2048 & 91.8(3) \\
\bottomrule
\end{tabular}